\documentclass[%
 reprint,
superscriptaddress,
groupedaddress,
unsortedaddress,
altaffillsymbol,
 amsmath,amssymb,
 aps,
]{revtex4-2}

\usepackage{appendix}
\usepackage{longtable}
\usepackage{graphicx}
\usepackage{dcolumn}
\usepackage{bm}
\usepackage{tablefootnote}
\usepackage{footnote}

\begin{document}

\preprint{APS/123-QED}

\title{Coexistence of single particle and collective excitation in $^{61}$Ni}

\author{Soumik Bhattacharya \thanks{corresponding author : soumik.kgpiit@gmail.com}, Vandana Tripathi \thanks {corresponding author : vtripathi@fsu.edu}, E.~Rubino, Samuel Ajayi, L.~T. Baby, C.~Benetti, R.~S.~Lubna, S.~L.~Tabor}
\affiliation{Department of Physics, Florida State University, Tallahassee, Florida, 32306, USA}
%
%

\author{J.~D\"oring} 
\affiliation{Bundesamt für Strahlenschutz, D-10318 Berlin, Germany}

\author{Y.~Utsuno}
\affiliation{Advanced Science Research Center, Japan Atomic Energy Agency, Tokai, Ibaraki 319-1195, Japan}
\affiliation{Center for Nuclear Study, University of Tokyo, Hongo, Bunkyo-ku, Tokyo 113-0033, Japan}


\author{N.~Shimizu}
\affiliation{ Center for Computational Sciences, University of Tsukuba,
1-1-1 Tennodai, Tsukuba 305-8577, Japan}

\author{J. M. Almond}
\affiliation{Physics Division, Oak Ridge National Laboratory, Oak Ridge, TN 37831, USA}

\author{G.~Mukherjee}
\affiliation{Variable Energy Cyclotron Centre, Department of Atomic Energy, 1/AF Salt Lake City, Kolkata-700064, India}


\date{\today}

\begin{abstract}

The high spin states in $^{61}$Ni have been studied using the 
 fusion evaporation reaction, $^{50}$Ti($^{14}$C,3n)$^{61}$Ni at an incident 
 beam energy of 40~MeV. A Compton suppressed multi-HPGe detector setup, 
 consisting of six Clover detectors and three single crystal HPGe detectors 
 was used to detect the de-exciting 
$\gamma$ rays from the excited states. The level scheme has been extended 
up to an excitation energy of 12.8 MeV and a tentative  $J^\pi$~=~35/2$^+$. 
The low-lying negative parity levels are found to be generated by single particle excitation 
within the $fp$ shell and also excitations to the $g_{9/2}$ orbitals as explained
well with shell model calculations using the GXPF1Br+$V_{MU}$(modified) interaction. 
Two rotational structure of regular E2 sequences with small 
to moderate axial deformation have 
been established at higher excitation energy. Most interestingly, 
two sequences of M1 transitions 
are reported for the first time and described as magnetic rotational bands. 
The shears mechanism 
for both the bands can be described satisfactorily by the geometrical model. 
The shell model calculation involving the cross shell excitation beyond the 
$fp$ shell well reproduce the M1 and E2 sequences. The shell model predicted 
B(M1) values for the magnetic rotational band B1 show the decreasing trend with spin
as expected with closing of the shears.

\end{abstract}

\maketitle


\section{Introduction}

Neutron-rich nuclei with A$\sim$60  have attracted the attention of experimental and 
theoretical studies in the last few decades because of the evolution of shell structure with
the $N/Z$ value in this region. On one hand, from the systematic studies of 2$_{1}^{+}$ energies 
and B(E2:2$_{1}^{+}$$\rightarrow$0$_{1}^{+}$) strengths, a new sub-shell closure has been 
established at $N=32$ 
in $^{52}$Ca~\cite{Gade06}, $^{54}$Ti~\cite{Jan02}, and $^{56}$Cr\cite{Pris01}. Contrary to 
that, the decreasing trend of the 2$_{1}^{+}$ and 4$_{1}^{+}$ energies towards $N=40$ 
\cite{Gade10, Adrich08}, for both the Fe and Cr isotopic chains 
point to the diminishing of the $N=40$ sub-shell closure for mid-shell nuclei between the 
Z=20 and 28 major shell-closures. For Ni isotopes, the $Z=28$ major shell closure stabilizes the
spherical shapes near the ground state between $N=28$ and $N=40$, and the low and moderate energy
states can be well reproduced by shell-model calculations involving the $\nu$p$_{3/2}$, 
$\nu$f$_{5/2}$, $\nu$p$_{1/2}$ and $\nu$g$_{9/2}$ orbitals~\cite{Broda12, Schiffer13}. Though the
N=40 subshell closure is visible in $^{68}$Ni~\cite{Bernas82}, the energy gap between the 
1p$_{1/2}$ and 0g$_{9/2}$ orbitals is estimated to be small to allow the cross-shell excitations 
in $^{68,67}$Ni which are well reproduced by shell-model calculations \cite{Broda12, Zhu12}. The
involvement of the shape driving $\nu$0g$_{9/2}$ orbital induces collectivity in 
lighter Ni isotopes,
and rotational bands have been observed in $^{56-59}$Ni~\cite{Johansson08, Rudolph10, Johansson09,
Yu02}. Furthermore, magnetic rotational bands and super-deformed bands have also been reported in
$^{60}$Ni\cite{Torres08Ni60} and $^{63}$Ni~\cite{Albers13Ni63}, respectively. Recently, excited 
states in $^{61}$Ni have been studied experimentally~\cite{Saradindu} via a $^{7}$Li induced 
reaction and the energy levels could be explained fairly well within the scope of shell-model 
calculation involving cross-shell excitations to 0g$_{9/2}$ single-particle orbital.

The odd-mass Ni isotopes with neutron number slightly over $N=28$ are  
ideal cases to address the
role of neutron g$_{ 9/2}$ excitations at high spin. In the investigation of
$^{59}$Ni ~\cite{Yu02} four rotational bands up to a probable spin of (43/2) with terminating 
properties were found.
Based on configuration-dependent cranked Nilsson-Strutinsky calculations, two bands 
maintained significant collectivity until the maximum spin was reached. In
$^{63}$Ni three collective bands up to spin and parity of ($57/2^{+}$ ) were identified 
~\cite{Albers13Ni63}. Model calculations of cranked Nilsson-Strutinsky type indicate 
that in $^{63}$Ni collective excitations
sustain at moderate and high spins. The shears mechanism in $^{60}$Ni has been 
described microscopically with the self-consistent tilted axis cranking relativistic mean-field 
theory~\cite{Zhao60Ni} and via geometrical model in Ref.~\cite{Sourav60Ni}. Along with the 
$^{60}$Ni\cite{Torres08Ni60}, the magnetic rotational band has also been reported in 
$^{62}$Co\cite{Niru62Co}. For both the nuclei, the occupation of proton hole in the 
high-j $\pi$f$_{7/2}$ and neutron particle in the $\nu$g$_{9/2}$ orbital are 
described to form the shear arms. Recently, an adiabatic and
configuration-fixed constrained triaxial CDFT calculations searched for the possible wobbling motions 
in $^{59-63}$Ni~\cite{HuNiWobb}. They predicted wobbling motion in $^{59,61,62}$Ni, 
specially for $^{62}$Ni they pointed out the Band D1 as the possible wobbling band 
involving $\nu$g$_{9/2}^2$ configuration. Being in the same Fermi level for proton 
and neutron, the intermediate odd-mass $^{61}$Ni is emerged as an optimistic ground 
to search for the similar excitations. For $^{61}$Ni isotope, collective
bands at high spins were not known according to the most recent study using the $^{59}$Co($^{7}$Li,
$\alpha$n) reaction in combination with a modern array of Ge detectors~\cite{Saradindu}. 

The aim of the present work is to populate high-spin states in
$^{61}$Ni by using a heavy-ion induced reaction, and to search for possible collective 
excitations at high spins. Since it is difficult to reach high spin states in $^{61}$Ni 
with stable target-beam combinations, we selected the combination of a $^{50}$Ti target 
and a long-lived radioactive $^{14}$C beam. $^{61}$Ni was produced via the 3n-evaporation 
channel with a very high yield. Excited states in $^{61}$Ni were studied previously using the
$^{58}$Fe($\alpha$, n) and $^{53}$Cr($^{11}$B, p2n) reactions~\cite{Wads77Ni61,Wads77Ni61-2}, 
and the $^{48}$Ca($^{18}$O, 5n) reaction~\cite{War78Ni61} where levels up to 5316 keV were 
identified, including a 799 keV transition depopulating the ($17/2^{+}$) level at 4818 keV 
to the 15/2$^+$ level at 4019 keV, which decays further mainly via a 593 keV transition to 
the 13/2$^{\neg}$ level at 3426 keV. In the most recent study~\cite{Saradindu}, using the 
reaction $^{59}$Co($^{7}$Li, $\alpha$n), states were identified up to a possible spin of 
17/2$^+$ at 6734 keV. In the current work, based on the coincidences observed, a complex 
level scheme has been constructed, including many newly observed transitions.

\section{Experiment}

Excited states of $^{61}$Ni were populated using the $^{50}$Ti( $^{14}$C,~3n) fusion-
evaporation reaction at 40 MeV incident energy at the John D. Fox Superconducting 
Accelerator Facility,  Florida State University (FSU). The beam energy was chosen to maximize 
the population of the 3n evaporation channel. An isotopically enriched (90$\%$) $^{50}$Ti foil 
of 500 $\mu$g/cm$^{2}$ thickness was used as  target. 
The decaying $\gamma$ rays from the excited states were detected using the FSU 
$\gamma$-ray array, consisting of six Compton-suppressed 
Clover HPGe detectors and three single-crystal detectors. Three Clover detectors were placed at
90$^{\circ}$, two Clovers are placed at 135$^{\circ}$ and one detector at 45$^{\circ}$ angles 
with respect to the beam direction. One single-crystal detector was placed at
90$^{\circ}$ relative to the beam axis and two single-crystal detectors were placed at 
135$^{\circ}$ with respect to beam direction. The pre-amplifier signals from the Clover detectors, 
single crystal HPGe, and BGO detectors were processed using a PIXIE based digital 
data acquisition system in 2 fold coincidence mode for the $\gamma$ - $\gamma$ coincidence measurements. 
The efficiency and energy calibrations of each detector were carried out using 
$^{133}$Ba and $^{152}$Eu standard radioactive sources, placed at the target position. 
The  calibration and efficiency beyond the 1408~keV transition of $^{152}$Eu source were 
done by the high-energy transitions from $^{56}$Co source made at FSU using a proton beam.

\section{Data analysis}
\begin{figure}[ht]
  \includegraphics[width=1.0\columnwidth,clip=true]{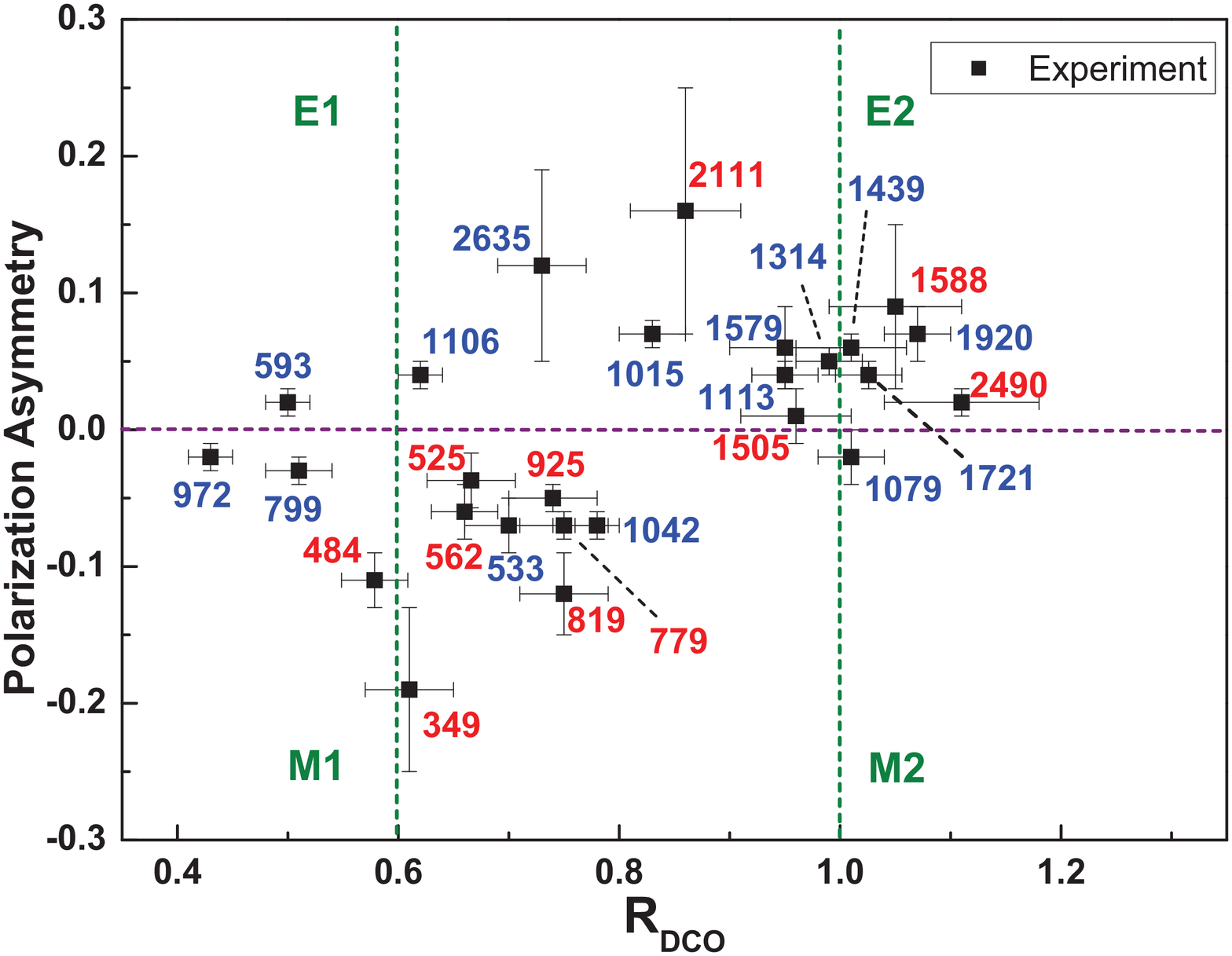}
\caption{\label{fig:DCOIP} DCO ratio vs polarization asymmetry ($\Delta_{IPDCO}$) of some of 
the known (marked blue) and new transitions (marked red) in $^{61}$Ni obtained from different 
quadrupole gates. The dotted lines (green) parallel to the Y-axis correspond to the R$_{DCO}$ 
values for dipole and quadrupole transitions in a pure quadrupole (E2) gate, respectively, and are 
shown to guide the eye. The dotted line (violet) parallel to the X-axis is to guide the eye for 
positive and negative values of $\Delta_{IPDCO}$ for determining the electric and magnetic 
transitions, respectively.}
\end{figure}

The time stamped data were sorted using GNUSCOPE, a spectroscopic software package developed at
FSU ~\cite{PavanGNU,GNU}.  A total of $9.4*10^{4}$ $\gamma-\gamma$ coincidence events have been 
accumulated from the present data. The event by event data were converted into a $\gamma-\gamma$ 
coincidence matrix with 1.0 keV/channel dispersion, which was also converted to a Radware~\cite{Radford} 
compatible matrix to analyze the coincidences between the de-exciting $\gamma$ rays. 
An asymmetric matrix was created using the data from the detectors at the backward (145$^{\circ}$) 
angles on the y-axis and the data from the 90$^{\circ}$ angle detectors on the x-axis to find out 
the Directional Correlation from Oriented states (DCO) ratio ~\cite{dco} for various transitions.
The Clover detectors at 90$^{\circ}$ were additionally  used for the 
measurement of Integrated Polarization from Directional Correlation of 
Oriented states (IPDCO)~\cite{Starosta,Droste} for assigning parity to the excited states.
The deduced DCO ratios and the IPDCO values were used to determine 
the multipolarities and the electric or magnetic nature of the transitions 
leading to J$^\pi$ assignments of the states wherever possible.

The multipolarities of the transitions belonging to $^{61}$Ni were obtained from the DCO ratios 
 (R$_{DCO}$), defined by
\begin{equation}
\label{eqn1}
R_{DCO} = \frac{I_{\gamma_1} \textrm{at $\theta$$_1$, gated by $\gamma$$_2$ at $\theta$$_2$}}{I_{\gamma_1} \textrm{at $\theta$$_2$, gated by $\gamma$$_2$ at $\theta$$_1$}} 
\end{equation}
The DCO ratio of an unknown ($\gamma_1$) transition is obtained from the ratio of its intensities 
at two angles $\theta_1$(145$^{\circ}$) and $\theta_2$(90$^{\circ}$) gated by another transition 
($\gamma_2$) of known multipolarity, as per the equation~\ref{eqn1}. For the present experimental 
set-up, the typical value
of R$_{DCO}$ for a known quadrupole or dipole transition (for $\gamma_1$) comes out to be 1.0 
when gated by a transition of the same multipolarity ($\gamma_2$). When gated by a
known stretched and pure (mixing ratio $\delta \sim 0$) quadrupole (dipole) transition ($\gamma_2$), 
then the R$_{DCO}$ value comes out to be close to 0.6 (1.7) for dipole (quadrupole) transition.

The parities of most of the excited states were determined by the polarization asymmetry measurement 
using the relative intensities of the parallel and perpendicular 
(with respect to the beam direction) Compton scattering of the emitted $\gamma$ rays, 
detected in the corresponding crystals of the Clover HPGe detector.  The 90$^{\circ}$ 
detectors are used for this purpose to maximize the sensitivity of the polarization 
measurements ($\Delta_{IPDCO}$) following the prescription of Ref.~\cite{Starosta, Droste}.

 \begin{figure*}
\centering
\includegraphics[width=0.95\textwidth]{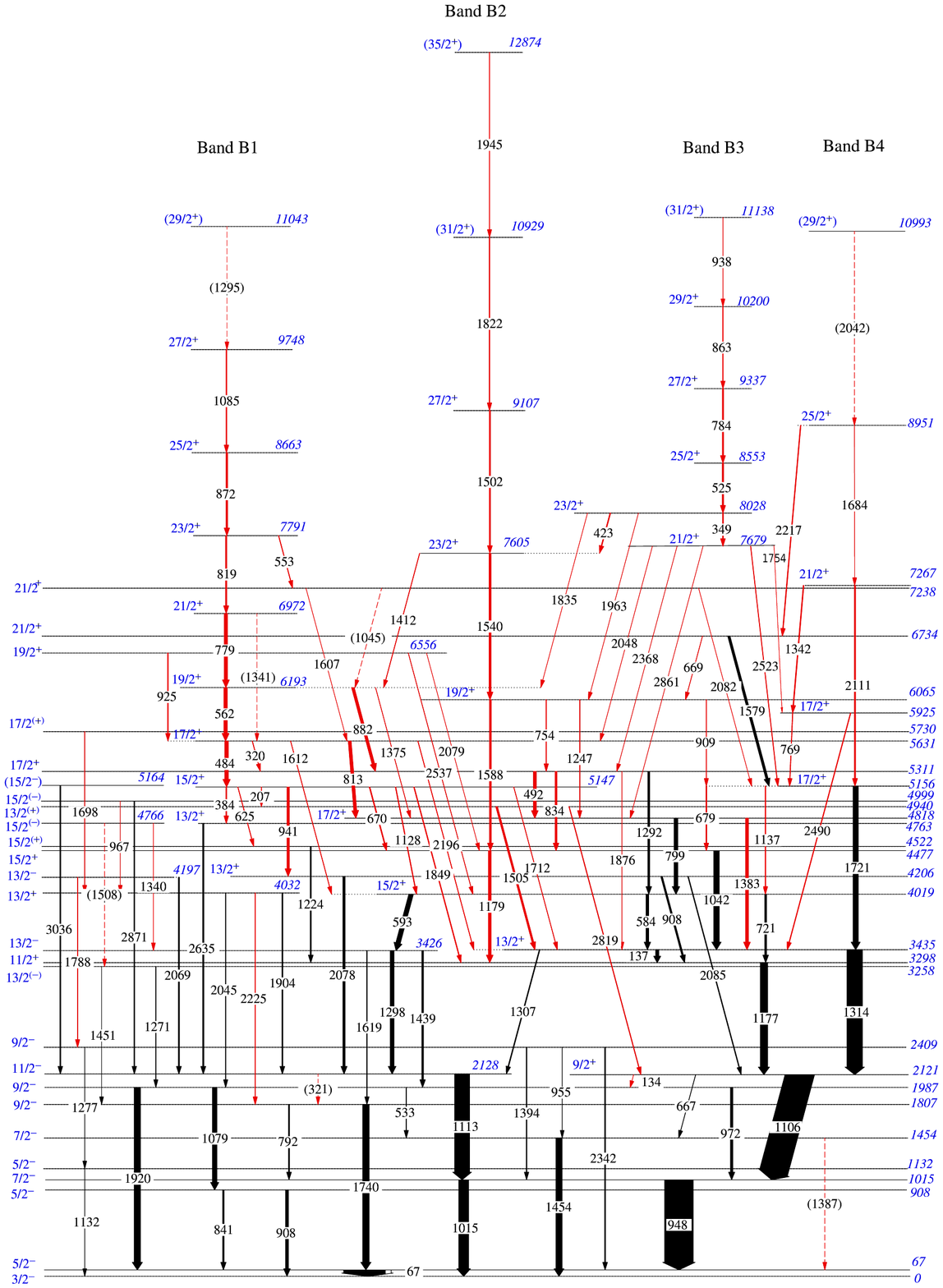}
\caption{\label{fig:LS}Level scheme for $^{61}$Ni from present work. The newly found transitions and
  repositioned transitions are shown in red. The width of the transitions
  represents the corresponding relative intensity.
  For clarity and easy comprehension of the level scheme, the energy values for the
  levels and the gamma rays are labeled to the nearest integer values.
  The accurate energy values  upto one decimal place are listed in Table.~\ref{tab:Table1} with corresponding error.}
\end{figure*}

The IPDCO asymmetry parameter($\Delta_{IPDCO}$) is defined as,
\begin{equation}\label{ipdco}
\Delta_{IPDCO} = \frac{a(E_{\gamma})N_{\perp} - N_{\parallel}}{a(E_{\gamma})N_{\perp} + N_{\parallel}}
\end{equation}
where N$_{\parallel}$ and N$_{\perp}$ are the total counts of the $\gamma$-ray, scattered events
in the planes parallel and perpendicular to the reaction plane, respectively. 
Here, a(E$_{\gamma}$) [= $\frac{N_{\parallel}}{N_{\perp}}$] is the geometrical correction factor 
(asymmetry factor) of the detector array and addresses the asymmetry in the response of the four 
crystals of a Clover Ge detector and was obtained from the known $\gamma$ rays from $^{152}$Eu 
unpolarized source as a function of $\gamma$-ray energy. The values of a(E$_{\gamma}$) for 
different $\gamma$-ray energies are estimated from the fitting of the data points from the 
unpolarized source with a linear equation [a(E$_{\gamma}$)=a$_0$+a$_1$E$_{\gamma}$] following the 
similar procedure described in Ref.~\cite{Soumik199Tl} whereas the fit parameter are found 
to be a$_0$=0.96385 and a$_1$=3.24*10$^{-5}$.
In order to calculate $\Delta_{IPDCO}$, the data were sorted to consider only the Compton events 
and two spectra were made with the total photopeak counts of parallel/perpendicular scattered 
events of the three 90$^{\circ}$ detectors. 
From these two parallel and perpendicular spectra, the number of counts 
in the perpendicular ($N_\perp$) and parallel ($N_\parallel$) scattering for a given $\gamma$-ray 
were obtained. The positive (negative) values of $\Delta_{IPDCO}$ correspond  
to the electric (magnetic) transitions are listed in Table ~\ref{tab:Table1} for various 
transitions in $^{61}$Ni. The DCO ratio and the $\Delta_{IPDCO}$ values
of various known and new transitions are also shown in Fig.~\ref{fig:DCOIP}.

\section{Results}

The level scheme of $^{61}$Ni deduced from the present coincidence measurement is shown in
Fig.~\ref{fig:LS} which is significantly extended up to a spin  of $35/2$
and excitation energy of 12874 keV with the 
observation of 76 new transitions and 28 new levels compared to the last published 
work~\cite{Saradindu}. All the low-lying known yrast and non-yrast states with the 
associated transitions, observed in the most recent high-spin study \cite{Saradindu} were 
confirmed with a few excpetions. At higher spin, a sequence of 483 (484~keV in Fig.~\ref{fig:LS}), 
562 and 941~keV M1 transitions were reported in Ref.~\cite{Saradindu}. The presence of those 
transitions could be verified but they have been placed differently in the level scheme 
Fig.~\ref{fig:LS}) according to the coincidence relationship from the present work.  
The present work reports the collective structure of $^{61}$Ni for the first time with 
the observation of four decay sequences referred  as Bands B1, B2, B3 and B4 in
Fig.~\ref{fig:LS} for the convenience of discussing them. Two of them, Bands B2 and B4 
are build with the transitions of quadrupole (E2 in this case) nature whereas 
in the Bands B1 and B3 the states are connected by strong M1 transitions. The level scheme 
is formed based on the different coincidence relationships observed and relative intensities 
among the transitions. The spin-parity of the levels are determined from the measured 
R$_{DCO}$ and $\Delta_{IPDCO}$ values of the transitions. 
The details of the levels, and the $\gamma$-ray transitions measured in this work, 
along with the DCO ratios and $\Delta_{IPDCO}$ values for all the transitions in $^{61}$Ni 
are tabulated in Table~\ref{tab:Table1}. 
\begin{figure}
\centering
\includegraphics[width=1.0\columnwidth,clip=true]{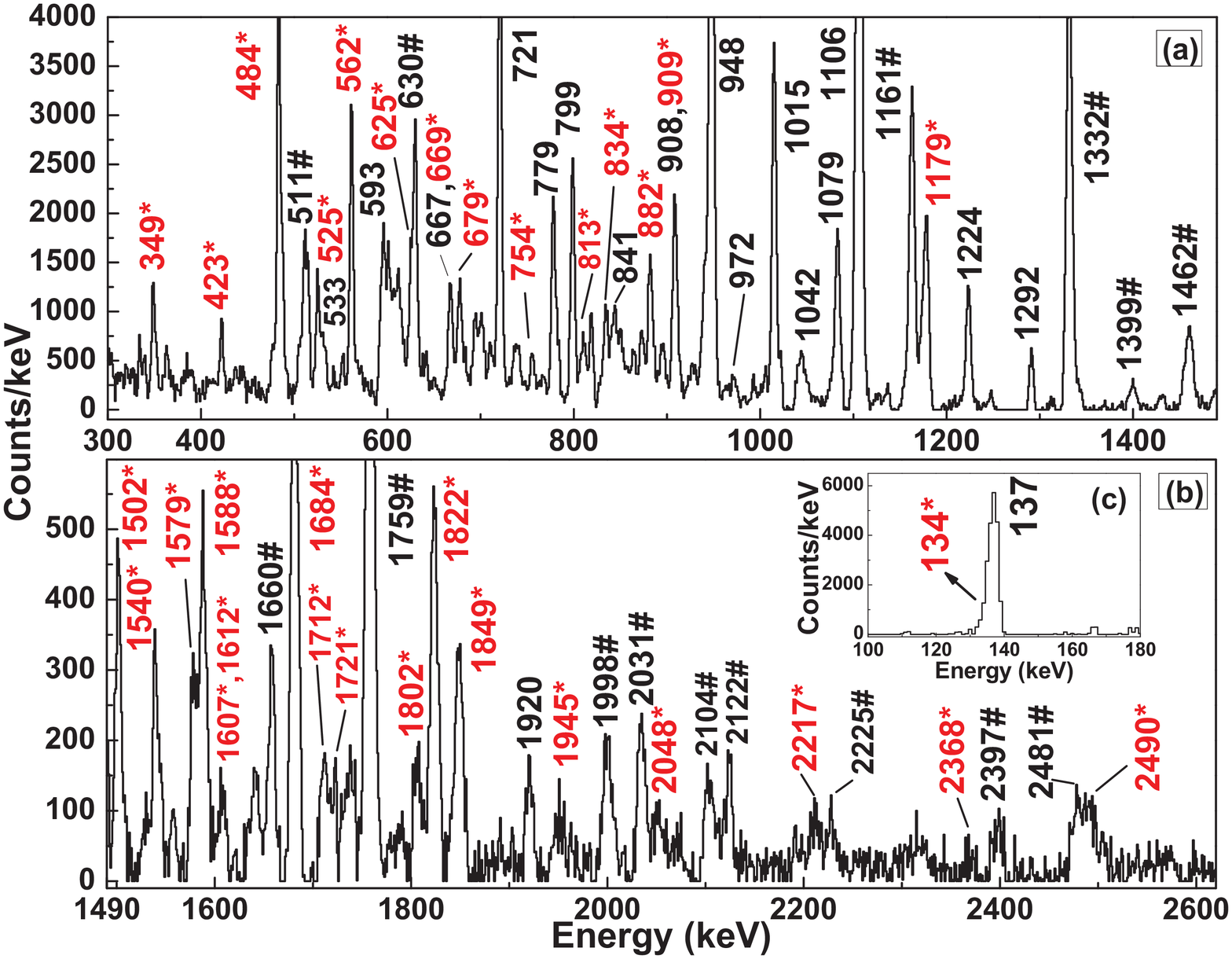}
\caption{\label{fig:g1177} Coincidence spectra gated by the 1177~keV transitions 
in $^{61}$Ni. The upper panel a (lower panel b) represents the $\gamma$ transitions 
from 300~keV to 1490 keV (1490~keV to 2620~keV) range. The inset panel (c) shows the presence of
 the 134 and 137~keV transitions.
`*' and red marked transitions are newly placed in the level scheme. 
The transitions marked with '\#' 
are the contamination from the neighbouring 
channels (mostly $^{60}$Ni).}
\end{figure}
The low-lying states, seen from the present work, are found to be of 
negative parity with a ground state of $J^\pi$ = $3/2^-$. 
The lowest opposite (positive) parity state $9/2^+$ is observed at 2121 keV. 
The 1177~keV $\gamma$ ray was reported to decay from  $11/2^{+}$ level at 3298~keV to the  
2121~keV $9/2^{+}$ spin state.  
The coincidence spectrum of the 1177~keV transition shown in Fig.~\ref{fig:g1177}
displays almost all $\gamma$-rays that were already known from prior work and are marked 
with black. Many new transitions are found to be present in this gate and marked with red 
(also *) in the figure. Interestingly, the presence of an 1179~keV peak in the 1177~keV gate establishes 
that it is a doublet with the 1179-keV transition de-exciting from the level at 4477 keV. 
 The absence of a 1314~keV transition in Fig.~\ref{fig:g1177} also confirms that the new 
 1179~keV transition is decaying to the 3298~keV level bypassing the 3435~keV level. 
The inset of Fig.~\ref{fig:g1177}(c) shows the presence of 134 and 137~keV 
transitions. The presence of 1079~keV and 972~keV peak also confirms the placement of the
134~keV transition from the 2121~keV $9/2^+$ state to the 1987~keV $9/2^-$ state.

\begin{figure}
\centering
\includegraphics[width=1.0\columnwidth,clip=true]{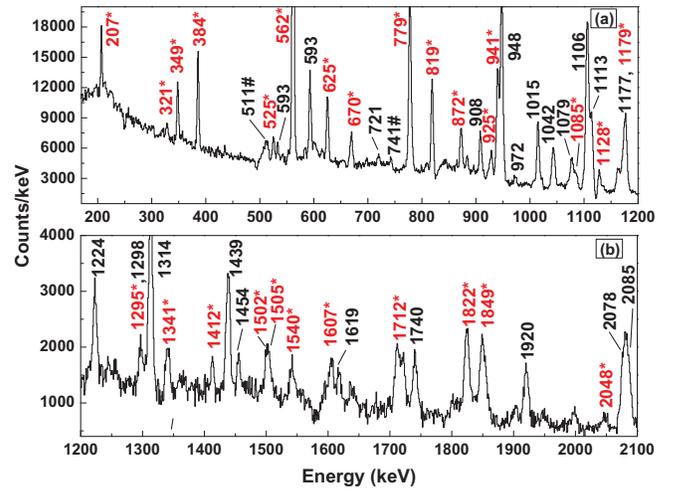}
\caption{\label{fig:g484} Coincidence spectra gated by the 484~keV transition corresponding to
band B1 in $^{61}$Ni. The upper panel a (lower panel b) represents the gamma transitions 
from 170~keV to 1200 keV (1200~keV to 2100~keV) range. All the M1 transitions corresponding to band 
B1 are shown along with the lower transitions.  
`*' and red marked transitions are newly placed in the level scheme.The transitions marked with '\#' 
are the contamination.}
\end{figure}

A representative coincidence spectrum for band B1 is shown in Fig.~\ref{fig:g484}. 
This band is found to be connected to the already reported 4477~keV and 4206~keV levels with 
670 and 941~keV connecting $\gamma$-transitions, respectively. The 384~keV transition is found to be
in coincidence with the already known 2635~keV transition decaying from 4763~keV state and thus placed
on the top of 4763~keV level.
The DCO ratio and $\Delta_{IPDCO}$ value of the 941~keV transition, as tabulated 
in Table~\ref{tab:Table1}, confirms 
it to be of M1 nature and thus the spin and parity of the 5147~keV state is fixed as $15/2^+$. 
The Band B1 consists of M1 transitions, namely 384~keV, 562~keV, 779~keV up to the weak 1295~keV 
and can be seen from the figure along with the other new and known transitions from $^{61}$Ni. 
The DCO ratios and $\Delta_{IPDCO}$ values of the transitions belonging to Band B1 are 
listed in Table~\ref{tab:Table1} which established the magnetic dipole nature of the transitions.
We have not identified any crossover E2 transitions for this band with the exception of 
one crossover transition, 1341~keV, that has been tentatively placed from the $21/2^+$ to 
the $17/2^+$ level. The spin-parity of the 5147~keV level is fixed as $15/2^+$ from the M1 nature 
of 941~keV transition which is decaying from 5147~keV state to the known 4206~keV $13/2^+$ state.

\begin{figure}
\centering
\includegraphics[width=1.0\columnwidth,clip=true]{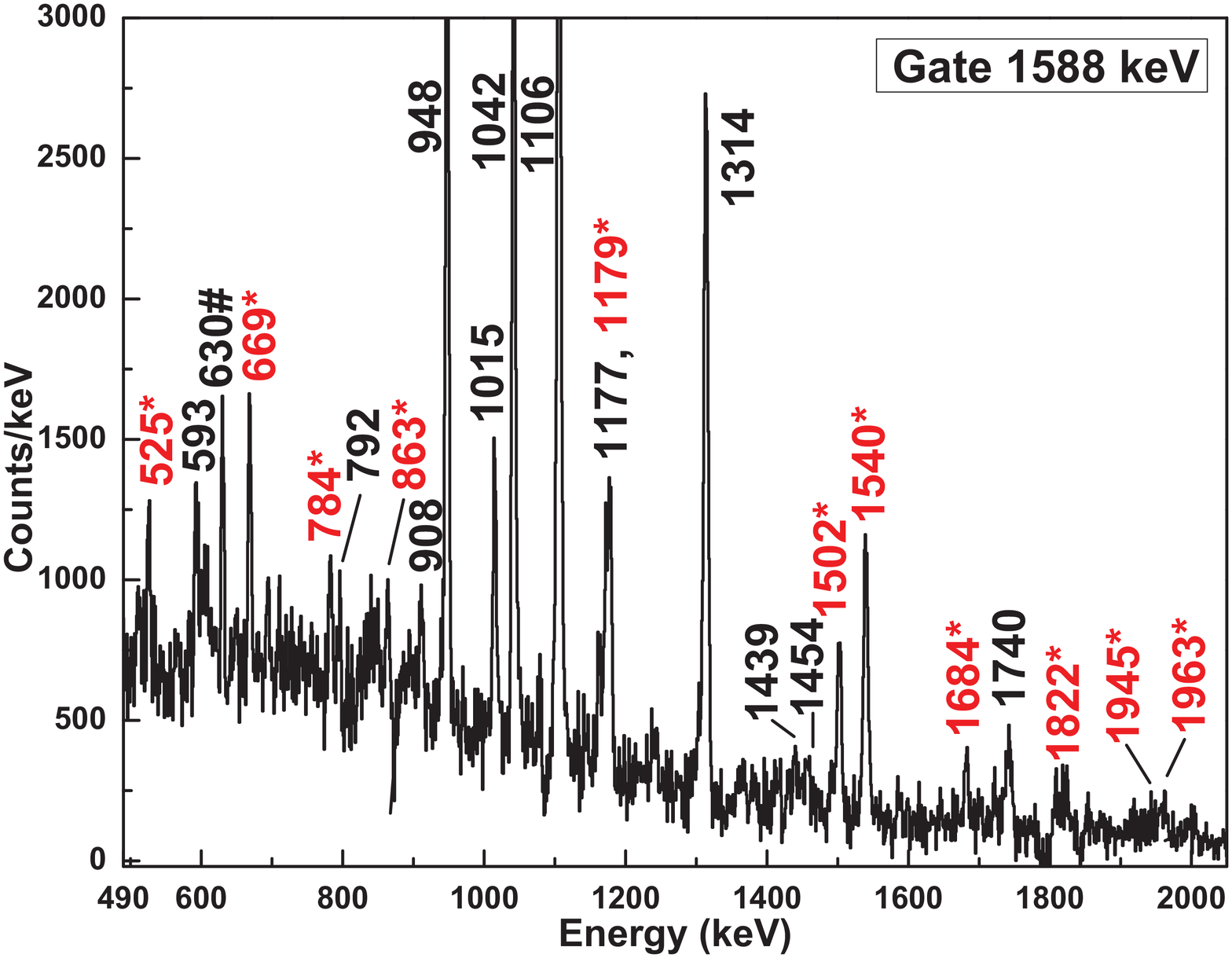}
\caption{\label{fig:g1588} Coincidence spectrum gated by 1588~keV showing the $\gamma$ transitions 
corresponding to the transitions of band B2 in $^{61}$Ni. `*' and red marked transitions are newly placed 
in the level scheme. The transitions marked with '\#' 
are the contamination from the weakly populated neighbouring channels.}
\end{figure}

The band B2 is made up of a sequence of E2 transitions. The newly found transitions 
corresponding to Band B2 
in coincidence with the 1588~keV (E2) transition are shown in Fig.~\ref{fig:g1588}. 
All the transitions (1179-, 1502-, 1540-, 1822- and 1945~keV) belonging to Band B2 can be 
seen in the 1588~keV gate.
This band is connected to Band B3 with the 1963~keV transition also seen in 1588~keV coincidence gate.
The regular pattern of E2 transitions starts from the 3298~keV level and extends up to 
the 12874~keV level with a plausible spin of $35/2^+$. The $R_{DCO}$ and $\Delta_{IPDCO}$ values confirm
the  spin and parity of the 4477~keV level  as $15/2^+$. The positive parity of the 
states in Band B2 is confirmed from the nature of the 1588 and  1502 keV transitions. 
The DCO ratio values could not be found for the 1822~keV and 1945 keV transitions due to limited
statistics, therefore the spin-parity of the 10929~keV and 12874~keV levels are only 
tentatively assigned.

\begin{figure}
\centering
\includegraphics[width=1.0\columnwidth,clip=true]{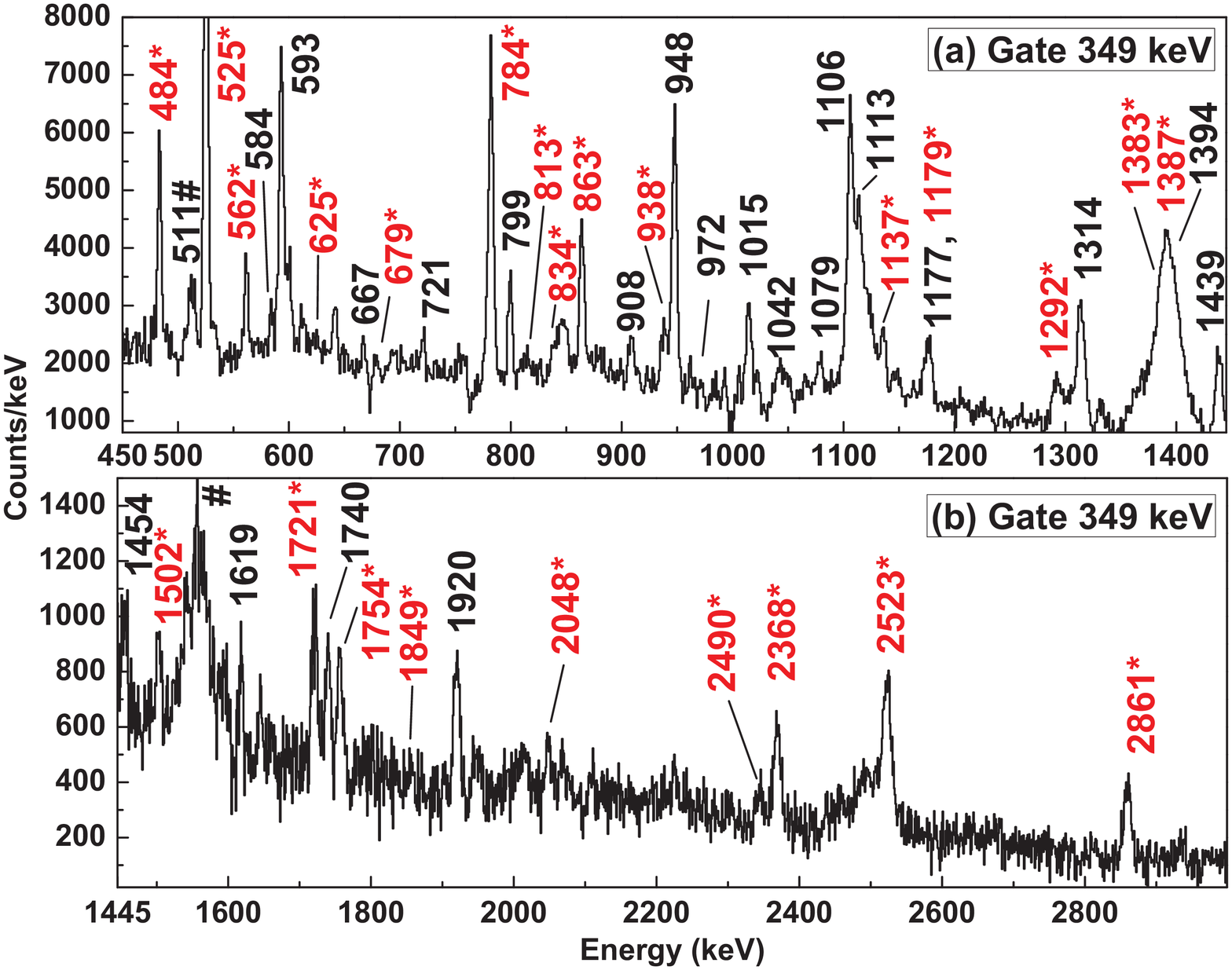}
\caption{\label{fig:g349} Coincidence spectra gated by 349~keV showing the (a) lower energy from 450~keV 
to 1445~keV and (b) higher energy from 1445~keV to 2999~keV region corresponding to the transitions of 
band B3 in $^{61}$Ni. `*' and red marked transitions are newly placed in the level scheme. The transitions
marked with '\#' 
are the contamination and the transitions from the neighbouring channels.}
\end{figure}

Another sequence of M1 transitions is found to be formed on top of the 7679~keV $21/2^+$ level. 
The spin of this level is fixed from the
quadruple nature of the 2523~keV transition which joins it to the
already known 5156~keV level. The corresponding new transitions of band B3 are shown in the coincidence 
gate of the 349~keV transition
in Fig.~\ref{fig:g349}. The presence of 2048, 2368, 2523 and 2861~keV transitions in the 349-keV gate 
confirms the connection of Band B3 with band B2 and different single particle structures. 
This band extends up to 11138~keV level with the sequence 
525, 784. 863 and 938~keV transitions on the top of the 349~keV transition, and these transitions 
show clear coincidences with the 349~keV transition (Fig. \ref{fig:g349}). 
The statistics corresponding to the 938~keV transition is limited and we could not
determine the $R_{DCO}$ or polarization asymmetry value for this transition. Thus the spin parity of 
the 11138~keV level is assigned as $31/2^+$ tentatively.  

\begin{figure}
\centering
\includegraphics[width=1.0\columnwidth, clip=true]{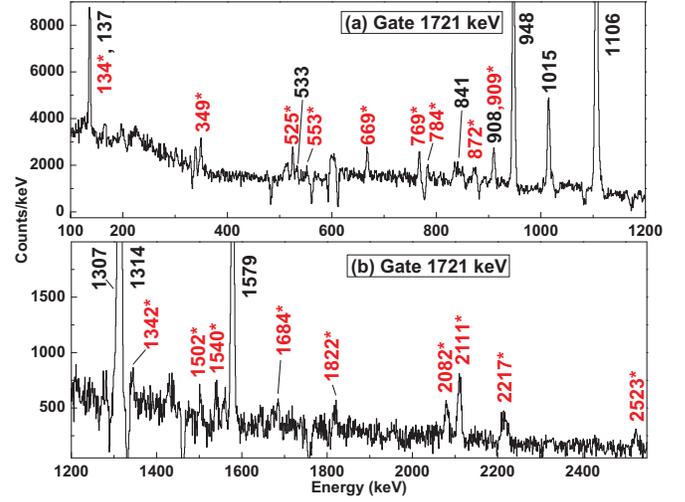}
\caption{\label{fig:g1721} Coincidence spectra gated by 1721~keV transition showing the 
(a) lower energy (100 to 1200~keV) and 
(b) higher energy (1200 to 2550~keV) transitions corresponding 
to the band B4 in $^{61}$Ni. 
`*' and red marked transitions are newly placed in the level scheme.}
\end{figure}

The Band B4 in the level scheme of $^{61}$Ni is populated rather weakly upto higher spin.
Two of its levels 3435 and 5156~keV were already known from the previous work connected through two E2
transitions 1721 and 1314~keV, respectively. The transitions belonging to Band B4 are 
confirmed by the coincidence with the  1721~keV transition, as
shown in Fig.~\ref{fig:g1721}. This band is found to continue up to $29/2^+$ 10993~keV level
with the sequence of 2111, 1684 and 2042~keV transition. The spin-parity of the $21/2^+$ 7267~keV level 
is confirmed from the E2 nature of 2111 ~keV level, whereas the spin of 8951~keV level is fixed as
$25/2$ from the $R_{DCO}$ value of 2217~keV connected to the previously known $21/2^+$ 6734~keV level.
Among the three newly placed 2111, 1684, 2042~keV transitions in Band B4,
the nature of the 2042 keV gamma cannot be confirmed because of overlapping with the already 
known 2045~keV transition, and thus the spin-parity of the 10993~keV level is tentatively assigned.
The spin and parity of the band-head of Band B4 is taken as the positive parity $9/2^+$
This E2 sequence is found to be weakly populated after spin $17/2^+$. 

\section{Discussion}
It is evident from the level scheme in Fig.~\ref{fig:LS} that with five valance neutrons outside the 
$^{56}$Ni doubly magic core, $^{61}$Ni is likely spherical at low spin and the lower excited 
irregular states are best described as single particle excitations. At higher spin, with more 
particle-hole excitations, collective behavior is expected and can be seen from several regular sequences
in the level scheme. 

\subsection{Low energy single particle structure and shell model calculations}
The present work confirmed the already established low energy excited states from the previous 
studies on $^{61}$Ni as well as established many new negative and positive parity states 
which decay to those well known states. Since Ni has a magic proton number (Z=28) and has only
five neutrons outside the N=28 shell, it is in the scope of the 
shell model calculations to interpret its low energy structures. To reproduce the experimental low 
energy states we used Shell Model (SM) calculations involving available orbitals near the Fermi surface. 
It has been seen that in the mass region Z or N from 20 to 40, natural parity states for odd(even) mass 
nuclei with $\pi$=-(+) can be described well within fp-shell model space with a inert core of $^{40}$Ca.
Therefore, for the low lying negative parity states, the 
shell model involves the 2p$_{3/2}$, 0f$_{5/2}$ and 2p$_{1/2}$ shells above the 
N, Z=28 gap and the 0f$_{7/2}$ shell below the gap.
However, to explain the unnatural parity state with $\pi$=+(-) for the odd (even) mass nuclei in 
this region, the model space within the major fp shell is not enough. Thus in the present calculation, 
to interpret the higher lying positive parity states in
odd mass $^{61}$Ni the inclusion of 0$\nu$g$_{9/2}$ and 1$\nu$d$_{5/2}$ is necessary along with 
the major fp shell. The present shell model calculation is 
carried out using the GXPF1Br+$V_{MU}$(modified) interaction and the model space is composed of  
(0f$_{7/2}$+2p$_{3/2}$+0f$_{5/2}$+2p$_{1/2}$)+ $\nu$g$_{9/2}$ + $\nu$d$_{5/2}$ 
orbits ~\cite{Togashi15}. The configuration space was truncated to allow up 
to six-particle excitations, from the 0f$_{7/2}$ shell to the upper fp-shell for protons. 
In the case of neutrons, to focus on the states dominated by the 1p-1h positive parity states 
across the N=40 shell gap, only one neutron is allowed to occupy the 0g$_{9/2}$ or 1d$_{5/2}$ orbitals.

\begin{figure}
\centering
\includegraphics[clip,width=\columnwidth]{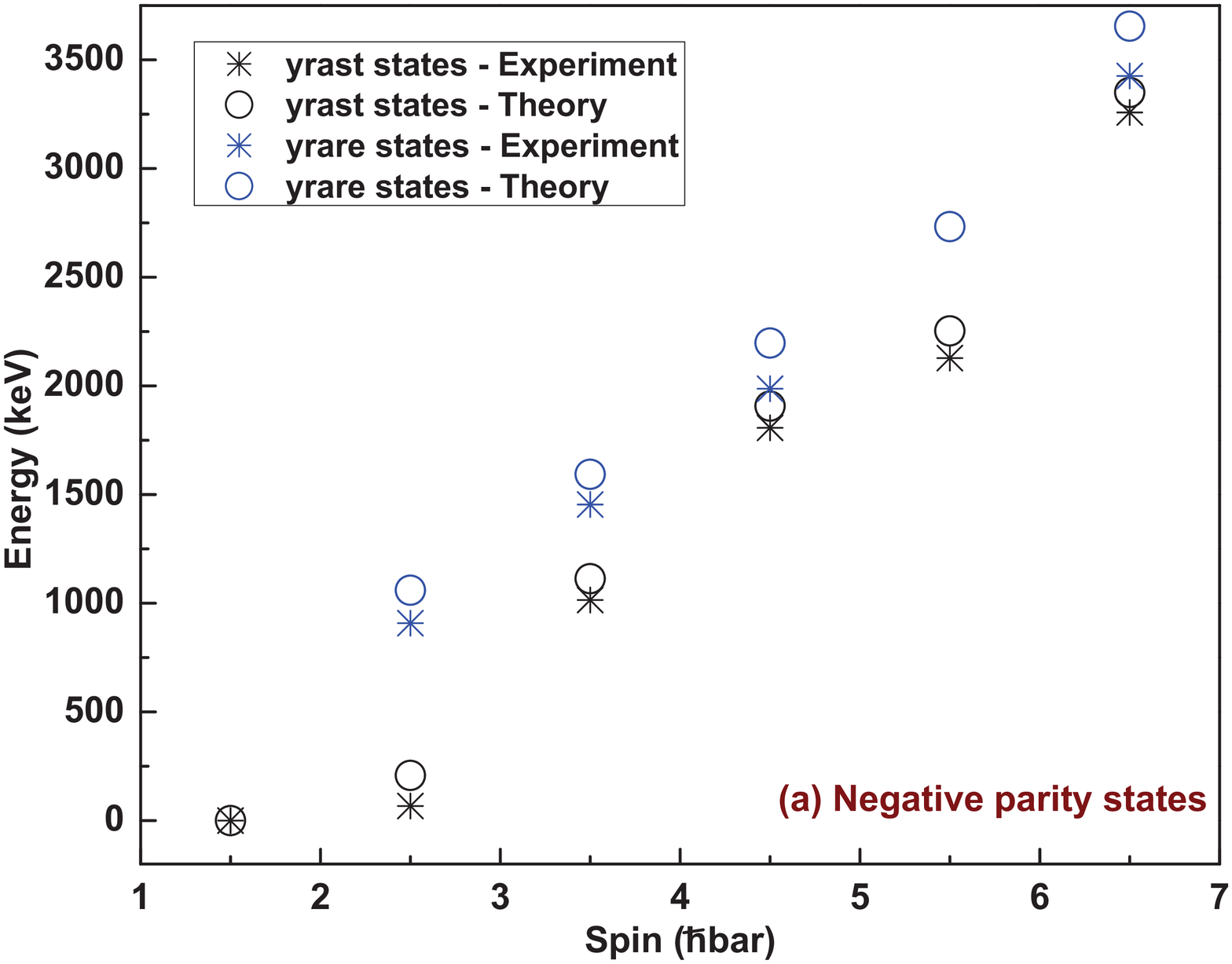}
\includegraphics[clip,width=\columnwidth]{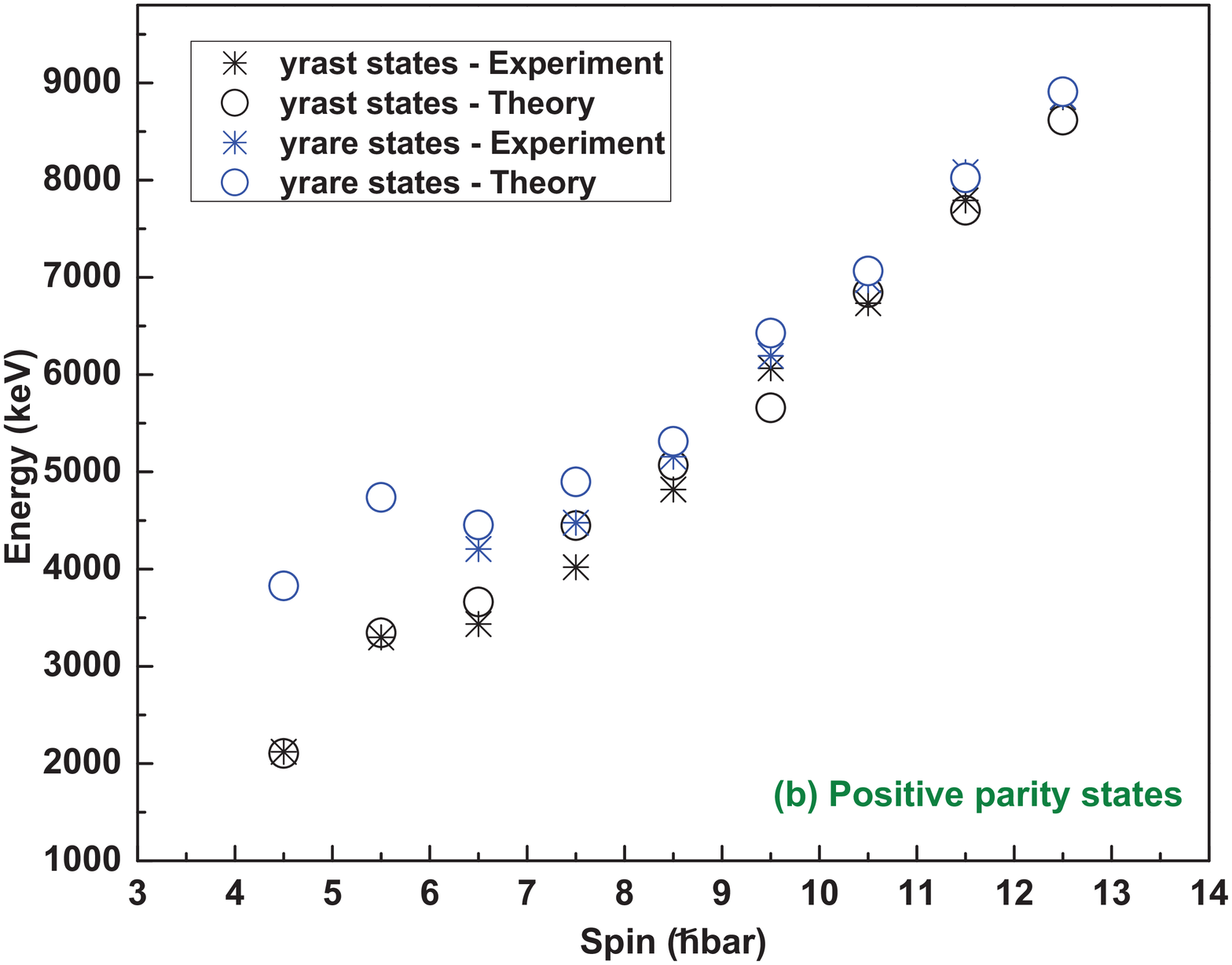}
\caption{\label{fig:SM_comp} The excitation energies of (a) negative parity and (b) positive parity 
low lying experimental states with SM calculations using the GXPF1Br+$V_{MU}$(modified) interaction. 
The first (yrast) and second excited (yrare) states for both experimental and theoretical levels of 
each spin are plotted. The experimental energy levels are marked with '*' whereas the SM calculated 
states are represented with open circles. The negative parity states are well reproduced using the 
$\nu$p$_{3/2}$, $\nu$f$_{5/2}$ and $\nu$p$_{1/2}$ orbitals, whereas for all the positive parity states, 
the neutrons are allowed to occupy the $\nu$g$_{9/2}$ or $\nu$d$_{5/2}$ orbitals.}
\end{figure}

The low-lying experimental negative parity and positive parity states are compared
with the SM generated states in Fig.~\ref{fig:SM_comp}. It is observed that both for 
positive parity and negative parity, the first experimental excited state of each spin 
that is the so called yrast states are matched quite well with the shell model predictions.
The SM generated negative parity states are predicted 
to have the proton mostly paired to spin 0 and the configuration of these states can be 
described as the odd neutron(s) occupying the $\nu$p$_{3/2}$, $\nu$f$_{5/2}$ or $\nu$p$_{1/2}$. 
The experimental yrare state for $1/2^-$ and $11/2^-$ spin have not been observed in 
the present data. 

On the other hand for the positive parity states, SM calculation shows that a neutron has 
to be excited into the positive parity $\nu$0g$_{9/2}$ orbital. 
The  first three SM predicted yrast states match reasonably 
well with the experimental yrast states. It may also be noted that we haven't 
seen any experimental yrare state for $9/2^+$ and $11/2^+$ spin.
Beyond those two spins, the yrast and yrare states are come closer  in energy as the spin 
increases. This is expected as with more energy and spin, more and more orbitals are 
accessible by the neutrons and protons for which the density of states are increasing 
significantly. Here it is worth mentioning that the yrast $9/2^+$ and $13/2^+$ states 
are part of the Band B4 and the SM calculation well reproduces those states of the Band 
B4 along with the other collective bands which will be discussed later.

\subsection{Bands B1 and B3}

\begin{figure}
\centering
\includegraphics[clip,width=1.0\columnwidth]{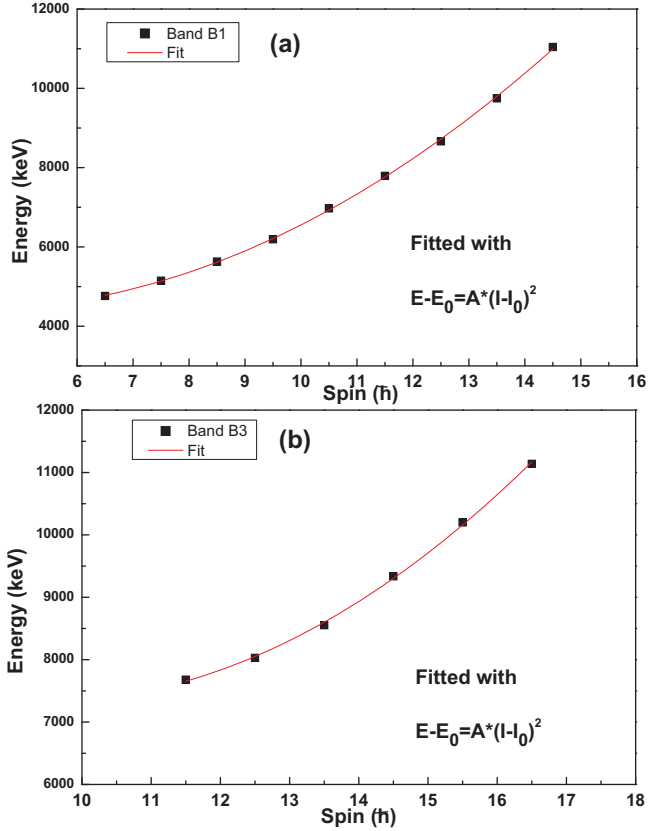}
\caption{\label{fig:BandFit_B1B3} The excitation energy of (a) Band B1 and (b) Band B3 as a 
function of the spin of the level. The experimental data are fitted with Eq.~\ref{BandFit} 
where $E_0$ and $I_0$ represent the band head energy and spin respectively.}
\end{figure}

In the present work we have found two regular sequence of M1 transitions, named
Bands B1 and B3, marked in Fig.~\ref{fig:LS}. 
Both are comprised of states with  positive parity. The involvement of the $\nu$0$g_{9/2}$ 
orbital plays a crucial role in the structure of these bands. To understand the 
nature of these bands, excitation energy as a function of spin has been plotted 
 in Fig~\ref{fig:BandFit_B1B3}. The excitation energy for both M1 sequences have 
 been fitted with
Eq.~\ref{BandFit} to understand their rotational nature.

\begin{equation}\label{BandFit}
E-E_0 = A * (I- I_0)^2 
\end{equation}

where $E_0$ and $I_0$ represents the band head energy and spin respectively and A is an 
arbitrary parameter.

It is evident from Fig~\ref{fig:BandFit_B1B3} that the Eq.~\ref{BandFit} 
fits the experimental data quite well for both the bands. Therefore, it can be said that 
the excitation energy of these two bands  follows the trend 
of a rotational band. For further investigation, these two sequences of 
magnetic dipole transitions have been studied in
the framework of SCM description of shears 
mechanism~\cite{Frn1993} prescribed by Machiavelli and 
Clark ~\cite{Clark2000, Mach98-1, Mach98-2}.

\subsubsection{Magnetic rotational band and shears mechanism}
Observation of rotational-like strong sequence of magnetic dipole (M1) 
transitions with very weak or no crossover E2 transitions in nearly spherical 
nuclei are often found to be generated from shears mechanism and well known
as magnetic rotational (MR) bands. This MR band structure is generated due to the 
symmetry breaking by the associated magnetic moments of
the current of few high-spin hole and particles outside a weakly deformed core. 
As these magnetic moments break the symmetry of the 
system and rotate around the total angular momentum of the near spherical core, 
this mode of nuclear excitations is described as magnetic rotational 
band \cite{Bald, Frn1993, Frn2001, Clark2000}.

\begin{figure}
\centering
\includegraphics[width=1.0\columnwidth]{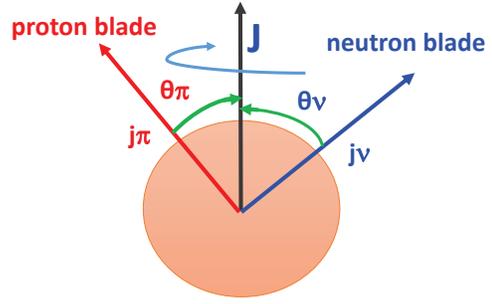}
\caption{\label{fig:Shear_mech}The orientation of proton (particle/hole) and 
neutron (hole/particle) angular momenta vectors with respect to the core angular 
momentum in the case of shears bands. With energy, the two angular momentum vectors 
(the arms of the shear closes down) align themselves with the core angular momentum 
due to interaction between the two angular momentum vectors.}
\end{figure}
\begin{figure}
\centering
\includegraphics[clip,width=1.0\columnwidth]{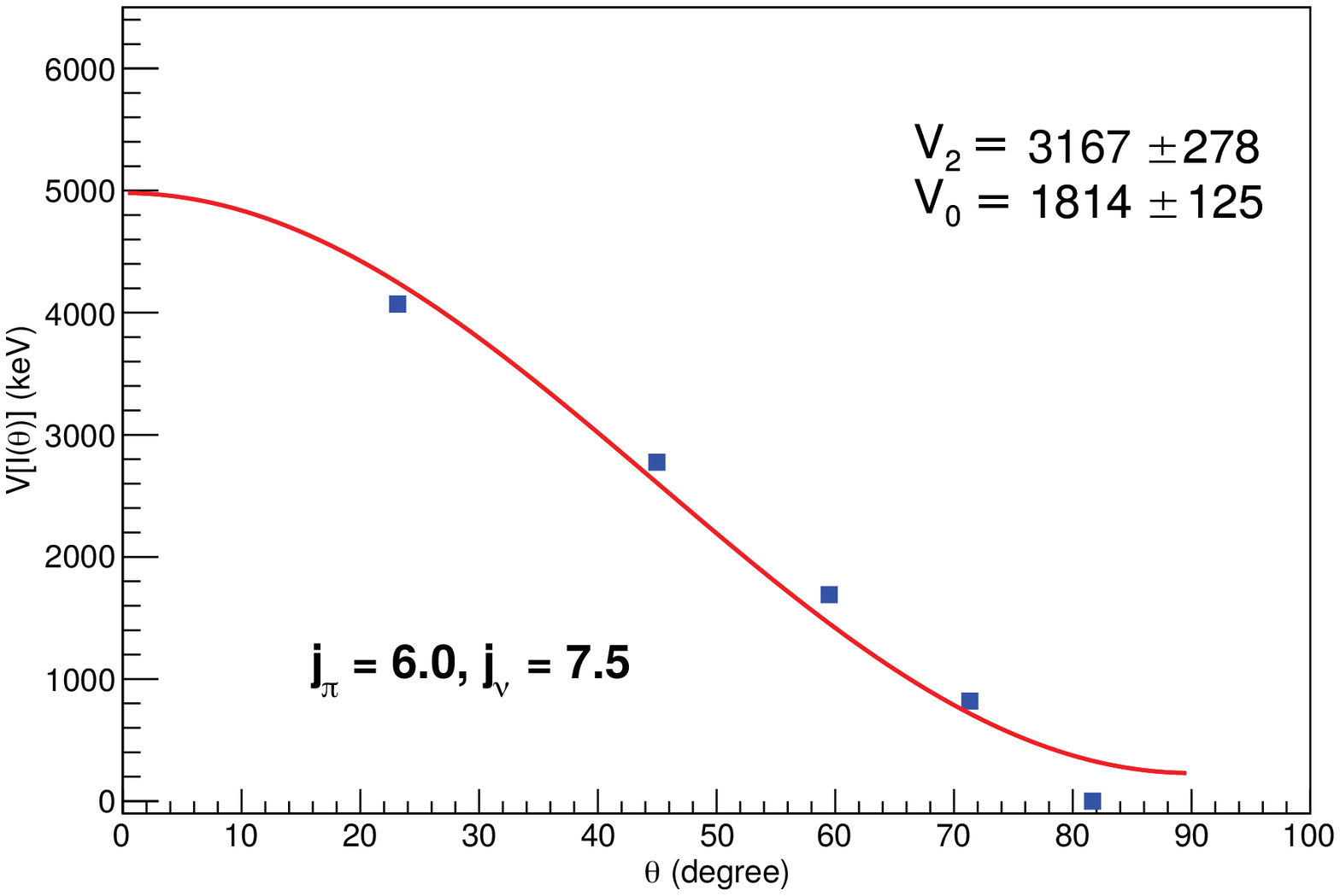}
\caption{\label{fig:B1_mag} SCM fit with the experimental data for magnetic rotational band B1.}
\end{figure}

\begin{figure}
\centering
\includegraphics[clip,width=1.0\columnwidth]{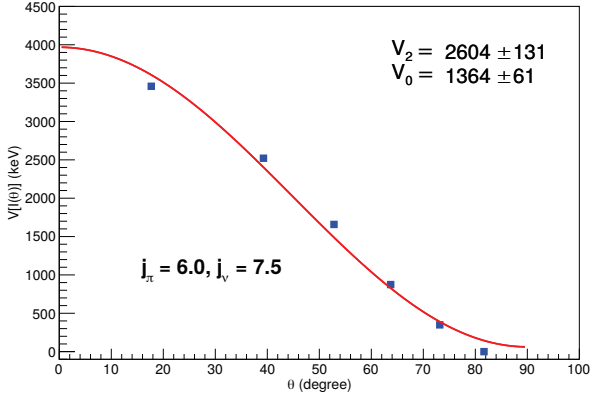}
\caption{\label{fig:B3_mag} SCM fit with the experimental data for magnetic rotational band B3.}
\end{figure}

 This type of sequence exhibits some special features like a strong B(M1)
 strength and a small B(E2) value, resulting in a large B(M1)/B(E2) value. The 
 current distributions of a few high spin particles and holes outside a near
 spherical core breaks the symmetry, resulting in these types of strong M1 
 sequences. The magnetic moments associated with these currents rotates around the 
 total angular momentum vector. 
 At the band head, the magnetic rotational band starts with a $90^{\circ}$
 angle between the two shear arms formed with particle and hole angular 
 momentum. With increasing excitation energy the two arms come closer and align themselves 
 with the angular momentum of the core, increasing the total angular momentum 
 of the system. At the highest point of the band the particle and hole angular 
 momentum is completely aligned and the angle between the two arms approaches zero.

The shear structure were first described by Frauendorf using the
tilted-axis-cranking model~\cite{Frn1993}, and a semi-classical model (SCM) was 
described by Macchiavelli et al.~\cite{Mach98-1}. The SCM describes 
schematically how the energy states of a shears band are generated from the
coupling of long spin vector of proton particles(or holes) $j_{\pi}$ and 
neutron holes(or particles) $j_{\nu}$. 
The shears angle $\theta (I)$ (= $\theta_{\pi}$ + $\theta_{\nu}$), 
i.e. the angle between two long spin vectors $j_{\pi}$ and 
$j_{\nu}$ as shown in Fig.~\ref{fig:Shear_mech}, is a function of total spin ($I$) and 
for a given  angular momentum state (of total spin $I$), it can be derived 
semiclassically using the expression
\begin{equation}
\label{MagAng}
cos[\theta (I)] = \frac { \overrightarrow {j_{\pi}} . \overrightarrow {j_{\nu}}} { |\overrightarrow {j_{\pi}}| .| \overrightarrow {j_{\nu}}|}=\frac{ I (I+1)-j_{\pi}(j_{\pi}+1)-j_{\nu}(j_{\nu}+1)} {2 \sqrt{(j_{\pi}(j_{\pi}+1)j_{\nu}(j_{\nu}+1))}} 
\end{equation}
The $j_{\pi}$ and $j_{\nu}$ are chosen to reproduce the band-head spin.
For a nucleus with small deformation, it has been observed that the total 
angular momentum has some contribution from the collective core in addition to that 
from the shears mechanism. Therefore, the total angular momentum
can be written as $\overrightarrow {I}=\overrightarrow {I_{Sh}}+ 
\overrightarrow {R_{core}}$ where the $R_{core}$ represent the core angular 
momentum. The small effect of the core towards the total angular momentum is
represented by,
\begin{equation}
\label{Rcore}
R_{core} = (\frac { \Delta R} {\Delta I }) * (I - I_{bh} ) 
\end{equation}
Using the band head spin $I_{bh}$, $j_{\pi}$, $j_{\nu}$ and the maximum spin 
observed in a band $I_{max}$ the $\Delta R$ and $\Delta I$ can be estimated as 
$\Delta R$ = $I_{max}$ - ($j_{\pi}$+$j_{\nu}$) and $\Delta I$=$I_{max}$-$I_{bh}$= $I_{max}$-$\sqrt{ {j_{\pi}}^2+{j_{\nu}}^2}$ which leads to 
\begin{equation}
\label{Rcore2}
\frac{\Delta R}{\Delta I} = \frac {(I_{max} - (j_{\pi}+j_{\nu}))} {(I_{max} - \sqrt{ {j_{\pi}}^2+{j_{\nu}}^2})} 
\end{equation}

Because of the effective interaction $V [I(\theta )]$ between the proton and neutron angular 
momenta, dynamics of the system gives rise to a rotation like spectrum consisting M1 transitions. 
The effective interaction $V [I(\theta) ]$ can be represented in terms of an expansion of even 
multipoles as stated in~\cite{Mach98-1}

\begin{equation}
\label{MagInt}
V[I(\theta)] = V_0 + V_2 \frac {3 cos^2\theta -1 } {2}+ ... 
\end{equation}
The excitation energies along the MR band are generated due to the effective
interaction by re-coupling of the two long angular momentum vector and can
be calculated from the experimental energy level of the band as $V[I(\theta)]$ = E(I) - $E_{bh}$.

With the aim to extract the effective interaction ($V_2$) between the nucleons 
involved in generation of angular momentum by the shears mechanism  for the 
different bands, we plot the E(I) - $E_{bh}$ (i.e $V[{I(\theta)}]$) vs the shears angle $\theta$(I) 
for band B1 and B3 using the above formalism and fit it with Eq.~\ref{MagInt} 
assuming the probable $j_{\pi}$ and $j_{\nu}$.

\begin{figure*}
\centering
\includegraphics[clip,width=1.9\columnwidth]{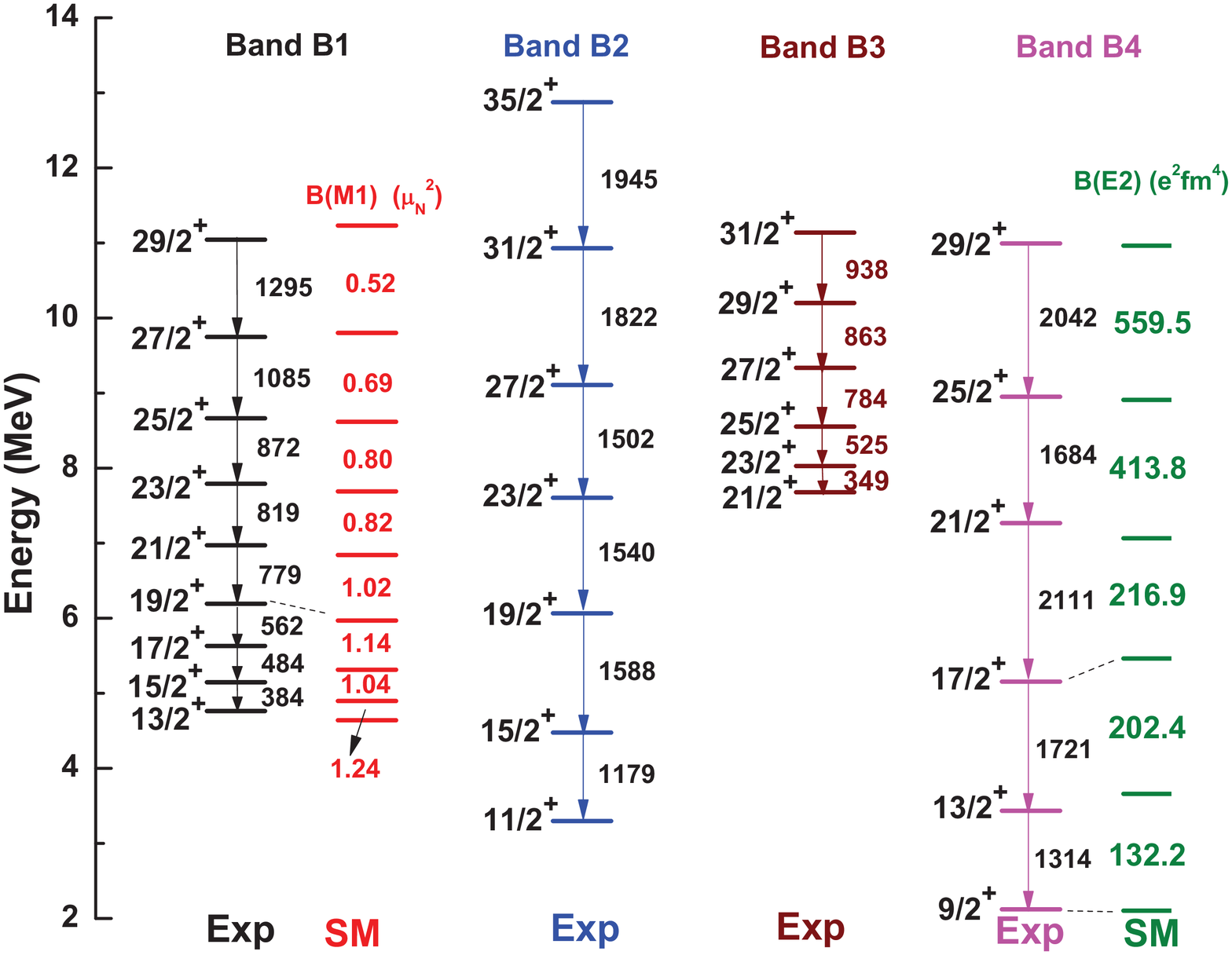}
\caption{\label{fig:SM_bandcomp} Simplified level structure of $^{61}$Ni showing  the four 
collective structures newly found in the present work. The shell model calculated levels 
(right panel) along with the experimental levels (left panel) are shown for Band B1 and B4 
for  comparison. The shell model calculation were done with the GXPF1Br+$V_{MU}$ (modified) 
interaction using the model space composed of  
(0f$_{7/2}$+2p$_{3/2}$+0f$_{5/2}$+2p$_{1/2}$)+ $\nu$g$_{9/2}$ + $\nu$d$_{5/2}$ orbitals both for 
proton and neutron with the 
truncation of 1p-1h states. The spins of each levels are marked for each band.
The dashed lines from the experimental levels to the SM calculated levels are to indicate 
corresponding spin states of each band. For each SM calculated level, the B(M1) values (for Band B1) 
and B(E2) values (for Band B4) are shown in the gap between the corresponding states.
For the experiment, the arrows indicate the $\gamma$ transitions and 
the energies (in keV) are also noted alongside.} 
\end{figure*}

In $^{60}$Ni the negative parity magnetic rotational bands 
(marked as M1 and M4 in ref.~\cite{Torres08Ni60}) are predicted to
have the configuration $\pi (f_{7/2}^{-1} (fp)^{1}) \otimes \nu (g_{9/2}^1 (fp)^3)$.
Considering the similar configuration for the dipole band B1 and B3 in $^{61}$Ni, 
the experimental data are fitted with the Eq.~\ref{MagInt}. 
The proton (hole) and 
neutron(particle) angular momentum ($j_\pi$ and $j_\nu$) 
corresponding to the blades of the shear are chosen so as to have the shears 
angle nearly $90^{\circ}$ at the band head energy as well as to generate the spin 
throughout the band by the gradual alignment of these two angular momenta. 
Depending on the configuration suggested in ref.~\cite{Torres08Ni60}, if we consider 
the $j_\pi$ and $j_\nu$ as 6$\hbar$ and 7.5$\hbar$ respectively for Band B1,
for the first few spin states upto $19/2^+ \hbar$ the shears angle is found 
to be more than 90$^{\circ}$. Thus the band head energy and the spin for Band B1 
is considered as 6972~keV and $21/2^+$ to plot the $V[I(\theta)]$ as a function of 
shears angle $\theta$ in Fig.~\ref{fig:B1_mag}. The SCM model fit using Eq.~\ref{MagInt}
matches quite well with the experimental data with the above considerations for 
the upper part of Band B1.
The effective interaction strength $V_2$=3167~keV from the fit in Fig.~\ref{fig:B1_mag} 
seems to match with the effective interaction of the magnetic rotational band 
reported in the neighbouring $^{60}$Ni~\cite{Sourav60Ni}. The shears angle is found to be 
about $90^{\circ}$ for the $21/2^+$ spin state and reaches about $25^{\circ}$ at the 
maximum observable spin ($29/2^+$) level for Band B1. 
For the first few states making up band B1, it seems there is less contribution from the 
$j_\pi$ and $j_\nu$ 
for the shears mechanism and therefore the geometrical 
model does not agree well for these states using $j_\pi=6 \hbar$ and $j_\nu= 7.5 \hbar$. 

\begin{figure}
\centering
\includegraphics[clip,width=1.0\columnwidth]{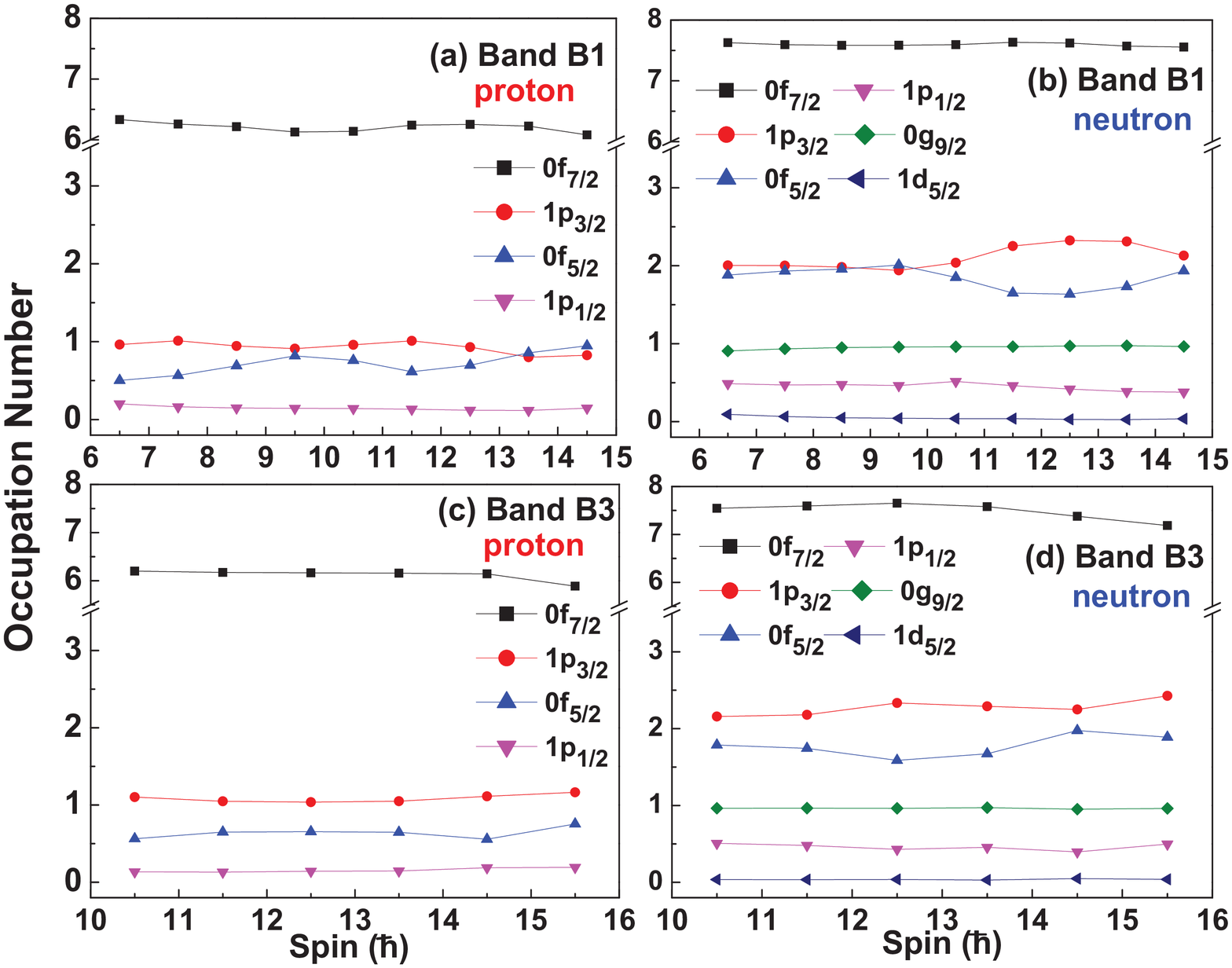}
\caption{\label{fig:B1B3_occ} The occupation number for different orbitals calculated by shell 
model calculation with respect to spins of the levels. The upper panel is consisting of the 
occupation number for (a) proton and (b) neutron associated with Band B1 configuration whereas 
the lower panel shows occupation number associated with the Band B3 configuration for (c) proton 
and (d) neutron.}
\end{figure}

For Band B3, same values of $j_\pi$ and $j_\nu$
are considered ($j_\pi$=6 $\hbar$ and $j_\nu$= 7.5 $\hbar$) to fit the experimental data.
The Eq.~\ref{MagInt} fitted the experimental data quite well for Band B3 as can be seen 
from Fig.~\ref{fig:B3_mag} considering the band head energy and spin of the Band B3 as 7679~keV and $21/2^+$ 
respectively. The interaction strength is found to be $V_2$=2604~keV 
which is a bit less compared to that of Band B1. It is seen from the 
Fig.~\ref{fig:B3_mag} that the shears angle $\theta$ is about $90^{\circ}$ at the band head
and gradually approaches $15^{\circ}$ at the highest spin.  
The Fermi surface of the proton–hole is near the $\pi$f$_{7/2}$ orbital in this nucleus and 
that for the neutron particles, it is near the $\nu g_{9/2}$ orbital. The proton holes 
in the high-j f$_{7/2}$ and neutron particle in the high-j $g_{9/2}$ orbital mainly build the two 
angular momenta arms of the shears bands discussed above. 
We find that the same value of $j_\pi$ and $j_\nu$ fits well for both the MR bands in $^{61}$Ni 
similar to what was observed for the configuration of the 
two negative parity magnetic rotational bands  (M1 and M4 in Ref.~\cite{Torres08Ni60}) in $^{60}$Ni. 
Further, We have tried to identify the corresponding states of Band B1 and B3 from the SM calculations
to understand their intrinsic structure as discussed in the next section.

\subsubsection{Magnetic dipole band and Shell Model Calculation}
The positive parity
states corresponding to the band B1 can be reproduced within the scope of the 
SM calculation as discussed before. The states which form the magnetic 
rotational bands are identified by their higher transition (B(M1) values)
probabilities associated with them. 
The experimental levels and Shell Model predicted states are compared in 
Fig.~\ref{fig:SM_bandcomp} for Band B1 and a good agreement is observed.
The comparatively large SM calculated B(M1) values are also shown in the corresponding 
gap of SM calculated states in Fig.~\ref{fig:SM_bandcomp}, which represent a nice 
decreasing trend with spin. Therefore, the important criteria of shears mechanism of having decreasing B(M1)
values across the band with the closing of two shears arms is well reproduced
with the SM calculation. To understand the configuration of the bands,
the occupation number of different orbitals for Band B1
(and also Band B3) are plotted as a function of spin and shown in the 
Fig.~\ref{fig:B1B3_occ}. For neutron, the occupation number for band B1 
clearly indicates that one neutron occupies the 
$g_{9/2}$ orbital whereas all the other four neutrons reside in the fp shell. 
If we follow the trend of the occupation number for $1 p_{3/2}$
and $0 f_{5/2}$ orbitals with spin, we can see that there is a gradual change in occupancy 
especially after $19/2^+$ spin which is consistent with the geometrical model interpretation of Band B1. 
It may happen that only after $19/2^+$ spin, the neutrons are completely aligned and then the shears arms
formed by the protons and neutrons start closing to generate the higher spin. 
The shell model picture also supports the choice of the band head at $21/2^+$ 
in Fig.~\ref{fig:B1_mag} and the high $j_\nu$ value as used in the geometrical model is also justified.
Along with that, one have to keep in mind that the expected (favourable) 
$90^{\circ}$ angle between the proton and neutron arm at the band head in the shears mechanism
is based on the argument that protons are in pure
hole states and neutrons are in pure particle states (or vice versa).
For $^{61}$Ni, on the other hand, proton angular momentum
is created by $\pi (f_{7/2}^{-1} f_{5/2}^{1})$ configuration, which is a
mixture of hole and particle state. This may change
the favorable angle between these two arms and the angle can be more 
than $90^{\circ}$ at the band head.
Another important note to be made about the occupation number is that it 
can have contribution from the pairing correlation. 
With both proton hole and neutron particle in high-j orbital, 
the ideal situation of creating
a shears band exists and the high B(M1) values from the shell model calculations strengthen 
the concept of a  magnetic rotational band. 
Therefore, from the occupation of protons and neutrons in 
different orbitals, the upper part of Band B1 are predicted to have the 
configuration $\pi (f_{7/2}^{-1} [p_{3/2}/f_{5/2}]^{1}) \otimes \nu 
(g_{9/2}^1 p_{3/2}^2 [f_{5/2}/ p_{1/2}]^2)$ and they agree with the 
configuration proposed in $^{60}$Ni Band M1 as well.

The SM predicted energies for the Band B3 are a little bit over-estimated compared 
to the experimental data and therefore are not show in 
Fig.~\ref{fig:SM_bandcomp}. 
To understand the structure of Band B3, the neutron occupation numbers for 
various orbitals are compared to Band B1 in Fig.\ref{fig:B1B3_occ}(b) and (c). 
It may be worth mentioning that unlike the Band B1, the occupation number for neutron for 
Band B3 remains same for the entire spin range and matches with the later part of Band B1.
Therefore, from the SM calculation, the configuration for band B3 is predicted to be similar 
as Band B1 if not same. This configuration of Band B3 also supports the choice of same $j_\pi$ and $j_\nu$ 
values corresponding to the geometrical semi-classical model fit with respect to Band B1, 
as discussed in previous section. 

\subsection{Bands B2 and B4}

\begin{figure}
\centering
\includegraphics[clip,width=\columnwidth]{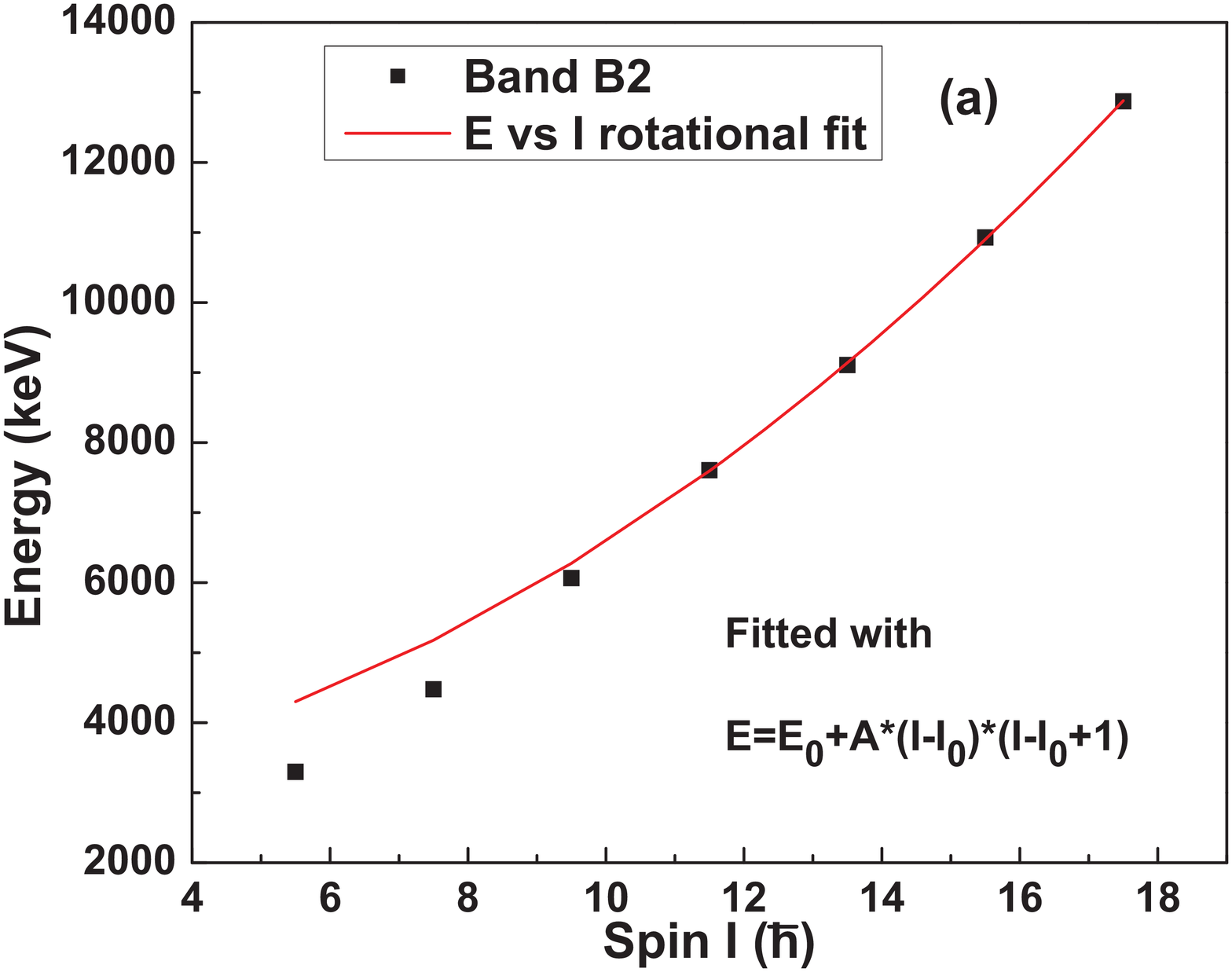}
\includegraphics[clip,width=\columnwidth]{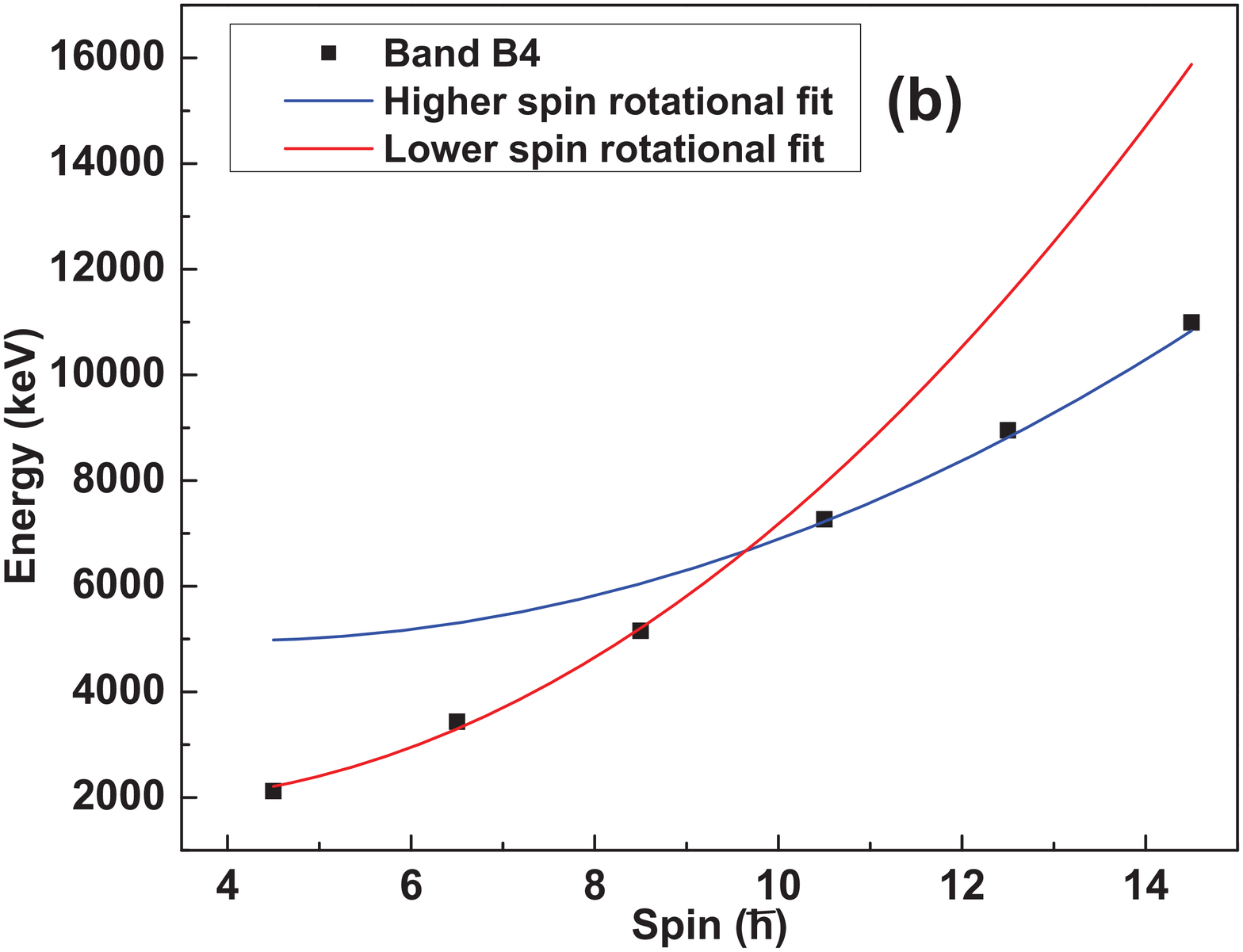}
\caption{\label{fig:BandFit_B2B4} The excitation energy of (a) Band B2  and (b) Band 
B4 as a function of the spin of the level. The experimental data are fitted with 
Eq. $E =E_0+ A* (I-I_0) (I-I_0+1)$, where A, $E_0$ and $I_0$ are varied as free parameter. 
$E_0$ and $I_0$ are equivalent to the band head energy and spin respectively for the 
corresponding Band (Band B2 and B4).}
\end{figure}

We have observed two more sequences connected by E2 transitions indicated as 
Band B2 and Band B4 in Fig.~\ref{fig:LS}. The Band B4 starts right from the first 
observed 9/2$^+$ spin and extends upto 
29/2$^+$ whereas the band B4 is assumed to start from 11/2$^+$ spin. 
It is quite clear that the involvement of the shape-driving g$_{9/2}$ orbital induces deformation and
the rotational bands are likely to be seen as also reported in neighbouring nuclei.
To find out the nature of these two bands we plot the excitation energy vs spin for 
both the bands (B2 and B4) which are displayed in Fig.~\ref{fig:BandFit_B2B4}(a) and (b).
The trend in the E vs I plot for Band B2 suggests that the 
first few transitions are not a part of the rotational structure,
whereas the transitions beyond the $23/2^+$ spin can be fitted with the 
rotational equation described as 
\begin{equation}\label{RotBandFit}
E = E_0 + A * (I- I_0)(I-I_0+1)
\end{equation}

\begin{figure}
\centering
\includegraphics[clip,width=1.0\columnwidth]{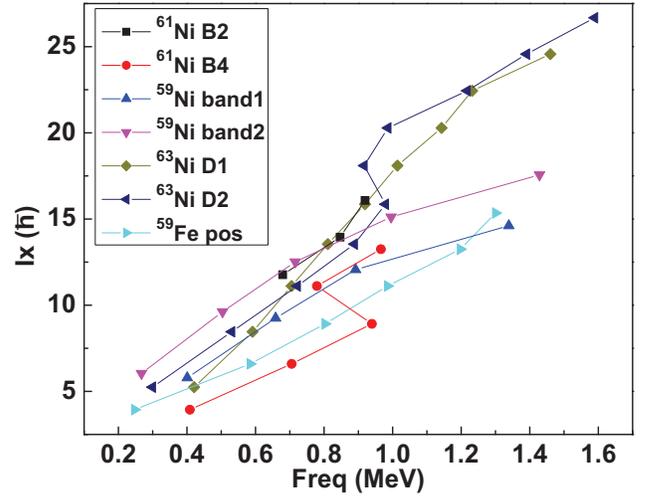}
\caption{\label{fig:allNi_Ix} The aligned angular momentum of Band B2 and Band B4 of $^{61}$Ni comparing with 
the rotational 
bands reported in neighbouring odd mass Ni isotopes and $^{59}$Fe as a function of the 
rotational frequency($\hbar \omega$) of the levels.
The information of the excited levels for $^{59}$Ni, $^{63}$Ni and $^{59}$Fe are
taken from the Ref.~\cite{Yu02},~\cite{Albers13Ni63} and~\cite{Deacon07Fe59} respectively.}
\end{figure}


The experimental data in Fig.~\ref{fig:BandFit_B2B4} for both the bands are fitted with 
Eq.~\ref{RotBandFit}. For Band B2, the fitting is for the upper 4 data points starting from 
$23/2^+ \hbar$ and clearly, this rotational structure is not followed by the states below $23/2^+$ spin. 
For band B4, the E vs I plot indicates the overlap of two sequences. Therefore they are fitted by the 
rotational equation in two part: lower energy part and higher energy part, which is shown 
in Fig.~\ref{fig:BandFit_B2B4}(b). The fitted curves indicate the crossing of 
two rotational bands around 23/2$^+$ spin.
The fitted parameter A in Eq.~\ref{RotBandFit} is inversely proportional to the Moment of 
Inertia (MoI) of the nuclear shape of associated with the corresponding band. 
The values obtained for A for Band B2 (upper part) (Fig.~\ref{fig:BandFit_B2B4}(a)) and 
Band B4 lower and upper part (Fig.~\ref{fig:BandFit_B2B4}(b)) are 28, 103 
and 53 respectively. Therefore, the higher slope of Band B4 compared to Band B2 indicates 
that a higher moment of inertia is associated with the Band B2 rotational structure. 
Thus the configuration for band B4 can be  assumed to have less number of 
quasi-particle  involving one $\nu 0g_{9/2}$ orbital and possibly
the upper part of Band B4 may have more quasi-particles.

To understand the structure better,
the aligned angular momentum (I$_x$) for bands B2 and B4, observed in $^{61}$Ni from 
the present work are compared with the similar rotational bands reported 
in neighbouring Ni and Fe nuclei and are shown in Fig.~\ref{fig:allNi_Ix}. 
As the Band B4 starts from the very first observed 9/2$^+$ state, the configuration 
$\pi (f_{7/2}^{-1} (fp)^{1}) \otimes \nu (g_{9/2}^1 (fp)^4 )$ is assumed for the lower 
part of the Band B4. As the aligned angular momentum is low for the first part of the band B4, 
we can assume that the few neutrons in fp shell are paired up and do not contribute to the 
total angular momentum. The aligned angular momentum (I$_x$) for the lower part of the Band 
B4 are matched with those of the positive parity band of $^{59}$Fe with only two proton less 
than $^{61}$Ni. Band B4 is seen to exhibit a back-bending at 0.95 MeV rotational frequency 
(around spin 10.5$\hbar$) and seems to have a different intrinsic structure after 21/2$^+$ spin.  
With only two more neutron compared to $^{59}$Ni, the rotational bands of $^{61}$Ni are 
expected to have similar kind of structure as the rotational bands reported in $^{59}$Ni. 
The configuration of rotational Band 1 of $^{59}$Ni is predicted to 
be $\pi (f_{7/2}^{-2} (p_{3/2}/f_{5/2})^{2}) \otimes \nu ((p_{3/2} f_{5/2})^2 g_{9/2}^1 )$ from 
configuration-dependent cranked Nilsson-Strutinsky (CNS) calculations~\cite{Yu02}. 
The Band B4 is not well extended after the band crossing, but the I$_x$ of the Band B4 after 
band crossing has the indication to match well with that of Band 1 of $^{59}$Ni and thus the 
upper part of Band B4 is proposed to be generated from promoting an 
extra proton to the upper fp shell and creating an extra hole in high-j f$_{7/2}$ 
which in turns increase the angular momentum of the system. 
The aligned angular momentum also supports the configuration of 
$\pi (f_{7/2}^{-2} (fp)^{2}) \otimes \nu (g_{9/2}^1 (fp)^4)$ for band B4 after the back-bending. 
The deformation for the configuration involving the $\nu g_{9/2}^1$ orbital in $^{61}$Ni 
is predicted to be $\beta_2$=0.24 from configuration-fixed constrained CDFT calculations in Ref.~\cite{HuNiWobb}. 
Therefore we can also assume the same deformation for the upper part of Band B4. The
configuration-dependent cranked Nilsson-Strutinsky (CNS) approach also predicts a similar 
deformation ($\epsilon_2$=0.22) for the collective band WD1 in $^{60}$Ni~\cite{Torres08Ni60} 
with the configuration  $\pi (f_{7/2}^{-2} (fp)^{2}) \otimes \nu (g_{9/2}^1) (fp)^3)$. 
Therefore the deformation of Band B4 is assumed to be of the same order as predicted in Ref.~\cite{HuNiWobb}. 

The I$_x$ for the upper part of Band B2 in $^{61}$Ni is only plotted in Fig.~\ref{fig:allNi_Ix} as the states
below 23/2$^+$ are not a part of the rotational structure. The I$_x$ of the Band B2 seems to have relatively 
higher values with respect to Band B4 and matched with the lower part of the D1 band of 
$^{63}$Ni~\cite{Albers13Ni63} as well as with band 2 of $^{59}$Ni. The positive parity Band D1 of $^{63}$Ni
is predicted to have the configuration involving  $\pi$g$_{9/2}$ and $\nu$g$_{9/2}^2$ to generate such large 
aligned angular momentum. Similarly, the band 2 of $^{59}$Ni is also predicted to have both $\pi$g$_{9/2}^2$ and 
$\nu$g$_{9/2}^1$ orbitals involved in the configuration. It may be noted that these bands in $^{59}$Ni or 
$^{63}$Ni appear at a higher excitation energy.
From the present experimental data, we could not extend the Band B2 beyond 35/2$^+$ and 
thus its difficult to predict the intrinsic structure for this band. 
But it may be inferred that the configuration related to Band B4 involve more high-j 
orbital(s) along with one $\nu$g$_{9/2}$ in it.

\subsubsection{Shell Model Calculations for Band 2 and 4} 

\begin{figure}
\centering
\includegraphics[clip,width=1.0\columnwidth]{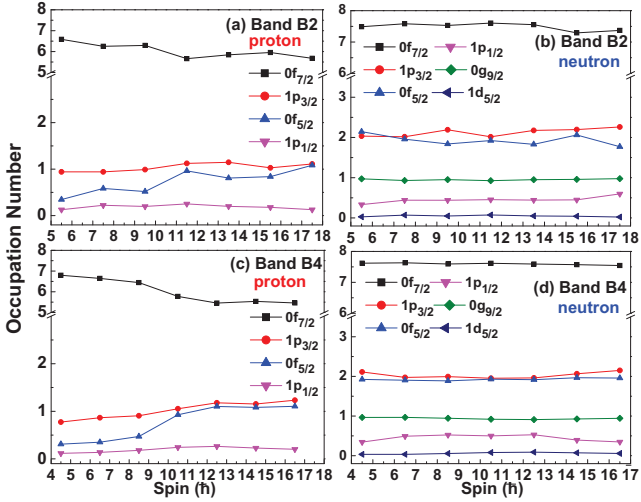}
\caption{\label{fig:B2B4_occ} The occupation number for different orbitals calculated by shell model 
calculation with respect to spins of the levels. The upper panel 
shows the occupation numbers for (a) proton and (b) neutron associated with Band B2 configuration 
whereas the lower panel shows occupation number associated with the Band B4 configuration 
for (c) proton and (d) neutron.}
\end{figure}

The experimental states corresponding to the rotational Band B4 are also compared with the 
SM calculated states and shown in Fig.~\ref{fig:SM_bandcomp}. The SM calculation
for Band B4 matched beautifully with the experimental value. This band is predicted to have a high B(E2) values
(in order of~200 e$^2$ fm$^4$) for the lower states which increase even higher after 21/2$^+$ spin.
The occupation number for both the Band B2 and B4 are represented in Fig.~\ref{fig:B2B4_occ}
calculated by the SM using same prescription described above. As these bands (Band B2 and B4) are associated 
with some deformation, the spherical orbital configuration are not so rigid. 
The occupation number for the 
Band B4 predicts one proton hole in f$_{7/2}$ and at least one neutron in shape driving g$_{9/2}$ orbitals 
for the lower spin with the most probable configuration of
$\pi (f_{7/2}^{-1} (p_{3/2})^{1}) \otimes \nu (g_{9/2}^1 (fp)^4$). As the band proceed to the higher 
spin it is clearly seen in Fig.~\ref{fig:B2B4_occ}(b), that the occupation of proton 
in $0f_{7/2}$ reduces after $21/2^+$ whereas the occupation in $0f_{5/2}$ increases. Therefore, an 
extra proton is predicted to be promoted in the fp shell from $0f_{7/2}$ after $21/2^+$ and the 
configuration of the higher spin states for Band B4
is emerged as $\pi (f_{7/2}^{-2} p_{3/2}^{1} f_{5/2}^{1}) \otimes \nu (g_{9/2}^1 (fp)^4)$. This
SM predicted configuration is well matched with the structure of band B4 discussed above. 
For Band B2 the SM predicted level energies for each spins are overestimated than the experimental 
states and thus are not shown along with experimental data in Fig.\ref{fig:SM_bandcomp}. 
However from the SM calculated occupation number of proton and neutron for Band B2 in 
Fig.~\ref{fig:B2B4_occ}(a) and (b), it may be predicted that this band also contains at least 
one proton hole in $0f_{7/2}$ and one neutron in $0g_{9/2}$.
The B(E2) values for this band is found out to be around 120 e$^2$fm$^2$ on average from SM. 
As discussed above, from the comparison with the neighboring nuclei in Fig.~\ref{fig:allNi_Ix}, 
the aligned angular momentum I$_x$ of this band is little bit higher and matched with the bands 
having configuration involving $\nu g_{9/2}^2$. It is beyond the scope of the present SM 
calculation to have more than one particle to cross the fp shell and occupy the $g_{9/2}$ 
or $d_{5/2}$ orbitals. That may be the reason behind the fact that the experimental excited 
states of Band B2 do not match well with the SM calculated states.

\section{Summary}

In the present work we have studied the structure of $^{61}$Ni using the $^{14}$C beam
from the John D Fox laboratory at FSU incident on a thin $^{50}$Ti target and using the 
FSU gamma array to detect the $\gamma$ rays. 
The spectroscopic information on $^{61}$Ni has been extended considerably up to
$\sim$13 MeV and $35/2^+$ spin/parity with the
establishment of 28 new levels and 76 new transitions decaying from those 
levels. The spin and parity of most of the states have been assigned from 
DCO ratios and $\Delta_{IPDCO}$ measurements of the $\gamma$ rays. 
With Z=28 for $^{61}$Ni, the low lying negative-parity states are 
found to be generated from single particle excitations of the odd 
neutrons in the $\nu p_{3/2}$, $\nu f_{5/2}$ and $\nu p_{1/2}$ orbitals as 
expected. 
For the positive  parity states, one neutron must be promoted 
to the next shell 0$g_{9/2}$ or 2$d_{5/2}$ orbitals. 
The shell model calculation using the GXPF1Br+$V_{MU}$(modified) 
interaction within a model space of fp-shell + $\nu$g$_{9/2}$ + 
$\nu$d$_{5/2}$ reproduced the negative as well as positive parity states 
pretty well. Apart from the low lying irregular structure, two magnetic
rotational (MR) bands Band B1 and Band B3 with a regular sequence of M1 transitions 
have been established. The shears mechanism associated with the MR band are described by the semi
classical model. With a single value of $j_\pi$ and $j_\nu$, the SCM cannot fit the Band
B1 entirely. Therefore the values of $j_\pi$ and $j_\nu$ are chosen wisely to have the shears
mechanism starts at $21/2^+$ (upper part of Band B1) with the shears angle$\sim$90$^{\circ}$ and 
continued up to maximum observed spin at $29/2^+$ by closing the shears angle 
between the proton and the neutron arms. With this consideration, and SCM fit matched pretty well 
with the experimental data. The experimental levels corresponding to Band B3 
are found to be associated with same $j_\pi$ and $j_\nu$ values from SCM fit to form the shears.
The shell model calculations predict the configuration for these bands 
involving one high-j $\pi f_{7/2}$ holes and one high-j $\nu g_{9/2}$ particle.
The shell model predicted a high B(M1)
values for these two band which further indicate them to be of magnetic rotational type.
With the involvement of the shape driving $\nu g_{9/2}$ orbital,
a small deformation is expected to be induced into the system. The formation of 
two deformed collective bands named Bands B2 and B4 with regular E2
transitions established the onset of deformation in this nuclei at higher energy. 
The excited energies of the collective bands are plotted as a function of spin 
for better understanding. The Band B4 are extended beyond the band crossing and an extra proton
is predicted to be excited from the lower shell to upper fp shell around $21/2^+$.
The SM predicts the energy levels of the Band B1 and B4 matched quite well with the experimental 
data within the considered model space whereas it overestimated the energies for Band B2 and Band B3. 
The occupation number of different orbitals for proton and neutrons, 
corresponding to different bands are plotted as a function of spin for these collective structures 
which indicate the configurations associated with them.
The Band B2 is seen not to be well developed from the present data and future experimental effort 
to extend this band would reveal additional information. Along with the extension of the 
quadrupole bands, the lifetime measurements of the levels of the magnetic rotational 
bands (B1 and B3) will be interesting to understand those structure better.

\section{Acknowledgement}
This work was supported by the U.S.
National Science Foundation under Grant Nos. PHY-2012522 (FSU) and U.S. Department of Energy, office of Science 
under Awards No. DE-AC05-00OR22725 (ORNL). Y. Utsuno  and N. Shimizu acknowledge KAKENHI grants (20K03981, 17K05433), 
“Priority Issue on post-K computer” (hp190160,
hp180179, hp170230) and “Program for Promoting Researches on the Supercomputer Fugaku” (hp200130, hp210165).
The authors are thankful to Mr. Shabir Dar, VECC, Kolkata, India for providing his semi-classical geometrical model 
code to fit the magnetic rotational band. We acknowledge the useful discussion with Dr. S. Bhattacharyya, VECC, Kolkata, India.



\providecommand{\noopsort}[1]{}\providecommand{\singleletter}[1]{#1}%

\section{Appendix}

\begin{longtable*}{|c|c|c|c|c|c|c|c|}
  \caption{\label{tab:Table1}The energies of the $\gamma$-ray transitions (E$_\gamma$), energies of the initial levels (E$_i$), the spins and parities of the initial ($J^{\pi}_i$) and final   ($J^{\pi}_f$) levels along with the relative intensities (I$_\gamma$), DCO ratios ($R_{DCO}$) and IPDCO values ($\Delta_{IPDCO}$) and the proposed multipolarities for
all observed transitions in $^{61}$Ni from the present work are listed. 
The $\gamma$ ray energies are determined from spectra displayed with 1.0 keV/Ch dispersion. The errors listed adjacent to each $\gamma$ transition represent the fitting error. 
The observed $\gamma$-ray intensities are estimated from prompt spectra and
normalized to 100 for the intensity of 67-keV $\gamma$-ray. 
The spin and parity of the states which cannot be determined from the present data are either presented in parentheses or adopted from the earlier work. 
The multipolarity and nature of few transitions, for which R$_{DCO}$ or $\Delta_{IPDCO}$ could not be determined or ambigous, are presented in parenthesis or fixed from the initial and final level spin-parity value.\\
$*$The intensity is less than 0.2$\%$ of the intensity of 67~keV.\\
$\#$The intensity could not be determined from present data due to overlap with nearby energies.\\
$\$$ DCO ratio in E2 gating transition. \\
$\&$ DCO ratio in E1 gating transition. \\
}\\

\hline

$E_{\gamma}$~~ & ~~ $J^{\pi}_i$$\rightarrow$ $J^{\pi}_f$~~ & ~~$E_{i}$~~& ~~$E_{f}$~~ & ~~ $I_{\gamma}$(Err)
~~~ &~~~ $R_{DCO}$~~ &~~ $\Delta_{IPDCO}$~ &~ Deduced \\
(in keV) &  & (in keV) & (in keV) & & (Err)  &  & Multipolarity \\
 
\hline
\endfirsthead
\multicolumn{7}{c}{Table~{\ref{tab:Table1}}: Continued...}\\
\hline

$E_{\gamma}$~~ & $J^{\pi}_i$$\rightarrow$ $J^{\pi}_f$~ & ~~$E_{i}$~~& ~~$E_{f}$~~ & $I_{\gamma}$(Err)
~~~ & ~~~ $R_{DCO}$~~ &~~ $\Delta_{IPDCO}$~ &~ Deduced \\
(in keV) &  & (in keV)& (in keV) & & (Err) &  & Multipolarity \\

\hline
\endhead
\endfoot
\endlastfoot


\hline
 
67.2(1)   &$	 5/2^{-}  \rightarrow	  3/2^{-} $&  67.2(1)    &  0.0(-)             &   100(3)     &  0.63(5)$^{\$}$    & -	            & M1 \\
134.2(1)  &$	 9/2^{+}  \rightarrow	  9/2^{-} $&  2121.0(1)  &  1987.0(1)          &   0.32(4)      &  -		     & -	    & E1 \\
136.6(1)  &$	13/2^{+}  \rightarrow	 11/2^{+} $&  3434.9(2)  &  3298.4(5)          &   7.52(23)      &  -		     & -	    & M1 \\
207.0(1)  &$	15/2^{+}  \rightarrow   13/2^{(+)}$&   5147.3(3) &   4939.8(2)         &   0.97(4)      &  -		     & -	    & (M1) \\
320.2(2)  &$	17/2^{+}  \rightarrow	 17/2^{+} $&   5630.9(3) &  5310.7(2)          &   0.77(3)      &  -		     & -  	    & M1 \\
321.0(3)  &$	11/2^{-}  \rightarrow	  9/2^{-} $&   2128.3(1) &  1807.4(1)          &   $<$ 0.2$^*$	&   -		     & -	    & M1 \\
348.8(1)  &$	23/2^{+}  \rightarrow	 21/2^{+} $&   8027.7(2) &  7678.9(2)          &   2.14(6)      &  0.61(4)$^{\$}$    & - 0.19(6)     & M1 \\
384.2(2)  &$	15/2^{+}  \rightarrow	 13/2^{+} $&   5147.3(3) &  4762.9(2)          &   1.73(6)      &  -		     & -	    & M1 \\
422.7(3)  &$	23/2^{+}  \rightarrow	 23/2^{+} $&   8027.7(2) &  2128.3(1)          &   1.71(4)      &  0.78(5)$^{\$}$    & -0.06(2)      & M1+E2 \\
483.7(1)  &$	17/2^{+}  \rightarrow	 15/2^{+} $&   5630.9(3) & 5147.3(3)           &   8.29(17)      &  0.58(3)$^{\$}$    & -0.11(2)      & M1 \\
492.4(1)  &$	17/2^{+}  \rightarrow	 17/2^{+} $&   5310.7(2) & 4818.3(2)           &   6.9(8)       &  -		     & -	    & M1 \\
525.0(2)  &$	25/2^{+}  \rightarrow	 23/2^{+} $&   8552.7(3) &  8027.7(2)          &   2.99(14)     &  0.67(4)$^{\$}$    & -0.06(2)      & M1 \\
532.8(2)  &$	 9/2^{-}  \rightarrow	  7/2^{-} $&   1987.0(1) &  1454.4(1)          &   1.06(2)      &  0.70(4)$^{\$}$    & -0.07(2)      & M1+E2 \\
552.6(1)  &$	23/2^{+}  \rightarrow	 21/2^{+} $&   7790.6(3) &  7237.9(3)          &   0.88(3)      &  1.00(5)$^{\&}$    & -0.25(15)     & M1 \\
562.0(1)  &$	19/2^{+}  \rightarrow	 17/2^{+} $&   6192.9(2) &  5630.9(3)          &   7.73(16)      &  0.66(3)$^{\$}$    & -0.06(2)      & M1 \\
583.8(1)  &$	15/2^{+}  \rightarrow	 13/2^{+} $&   4018.7(2) &  3434.9(2)          &   6.30(13)      &  0.95(5)$^{\$}$    & -0.20(5)      & M1+E2 \\
592.8(1)  &$	15/2^{+}  \rightarrow	 13/2^{-} $&   4018.7(2) &  3426.0(2)          &   10.56(25)    &  0.50(2)$^{\$}$    & 0.02(1)	    & E1 \\
625.2(2)  &$	15/2^{+}  \rightarrow  15/2^{(+)} $&   5147.3(3) &   4522.0(5)         &   0.72(3)      &  0.90(7)$^{\&}$    & - 	            & (M1+E2) \\
666.8(3)  &$	 9/2^{+}  \rightarrow	  7/2^{-} $&   2121.0(1) &  1454.4(1)          &   0.34(2)      &  0.59(7)$^{\$}$    & -		    & E1 \\
669.2(3)  &$	21/2^{+}  \rightarrow	 19/2^{+} $&   6734.3(2) &  6065.1(2)          &   $<$ 0.2$^*$  &  -		     & -	    & M1 \\
670.4(3)  &$	15/2^{+}  \rightarrow    15/2^{+} $&   5147.3(3) &  4476.9(2)          &   1.85(5)      &  0.94(7)$^{\$}$    & - 	            & M1+E2 \\
678.8(2)  &$	17/2^{+}  \rightarrow    15/2^{+} $&   5155.7(2) &  4476.9(2)          &   1.83(5)      &  1.10(6)$^{\&}$    & - 	            & M1+E2 \\
720.6(2)  &$	15/2^{+}  \rightarrow	 11/2^{+} $&   4018.7(2) & 3298.4(5)           &   4.23(9)      &  1.54(8)$^{\&}$    & 0.12(4)	    & E2 \\
754.4(1)  &$	19/2^{+}  \rightarrow	 17/2^{+} $&   6065.1(2) & 5310.7(2)           &   1.28(4)      &  1.20(6)$^{\&}$    & -0.07(3)      & M1+E2 \\
769.2(6)  &$	17/2^{+}  \rightarrow	 17/2^{+} $&   5925.2(2) &  5155.7(2)          &   0.90(3)      &  0.85(6)$^{\$}$    & - 	    & M1+E2 \\
778.8(2)  &$	21/2^{+}  \rightarrow	 19/2^{+} $&   6971.6(3) &  6192.9(2)          &   7.31(19)     &  0.75(4)$^{\$}$    & -0.07(1)      & M1+E2 \\
783.9(2)  &$	27/2^{+}  \rightarrow	 25/2^{+} $&   9336.6(4) &  8552.7(3)          &   2.67(18)     &  0.73(4)$^{\$}$    & - 	    & M1+E2 \\
792.4(1)  &$	 9/2^{-}  \rightarrow	  7/2^{-} $&   1807.4(1) &  1014.9(1)          &   2.18(6)      &  -		     & -	    & M1 \\
799.4(1)  &$	17/2^{+}  \rightarrow	 15/2^{+} $&   4818.3(2) & 4018.7(2)           &   8.63(18)      &  0.51(3)$^{\$}$    & -0.03(1)      & M1 \\
812.6(2)  &$	17/2^{+}  \rightarrow	 17/2^{+} $&   5630.9(3) & 4818.3(2)           &   6.9(7)       &  -		     & -	    & M1 \\
819.0(1)  &$	23/2^{+}  \rightarrow	 21/2^{+} $&   7790.6(3) & 6971.6(3)           &   2.93(7)      &  0.75(4)$^{\$}$    & -0.12(3)      & M1+E2 \\
833.8(1)  &$	17/2^{+}  \rightarrow    15/2^{+} $&   5310.7(2) &  4476.9(2)          &   4.31(14)     &  0.74(4)$^{\$}$    & - 	            & M1+E2 \\
840.6(1)  &$	 5/2^{-}  \rightarrow	  5/2^{-} $&	907.6(1) &    67.2(1)          &   3.54(32)      &  -		     & -	    & M1 \\
863.4(2)  &$	29/2^{+}  \rightarrow	 27/2^{+} $&  10200.0(4) &  9336.6(4)          &   1.12(8)      &  0.54(3)$^{\$}$    & - 	    & M1 \\
872.4(2)  &$	25/2^{+}  \rightarrow	 23/2^{+} $&   8663.0(3) &  7790.6(3)          &   3.59(9)      &  1.07(6)$^{\&}$    & -0.09(2)      & M1 \\
882.2(1)  &$	19/2^{+}  \rightarrow	 17/2^{+} $&   6192.9(2) & 5310.7(2)           &   6.12(13)      &  0.99(5)$^{\&}$    & -0.07(1)      & M1 \\
907.6(3)  &$	13/2^{+}  \rightarrow	 11/2^{+} $&   4206.3(2) &  3298.4(5)          &   3.02(10)      &  1.14(8)$^{\&}$    & - 	    & M1+E2 \\
907.7(1)  &$	 5/2^{-}  \rightarrow	  3/2^{-} $&	907.6(1) &  0.0(-)             &   6.9(7)       &  0.58(4)$^{\$}$    & - 	    & M1 \\
909.4(1)  &$	19/2^{+}  \rightarrow	 17/2^{+} $&   6065.1(2) &   5155.7(2)         &   0.69(8)      &  0.73(7)$^{\$}$    & - 	    & M1+E2 \\
924.7(3)  &$   19/2^{+}  \rightarrow	 17/2^{+} $&   6555.7(3) & 5630.9(3)           &   1.75(7)      &  0.74(4)$^{\$}$    & -0.05(1)      & M1+E2 \\
938.4(5)  &$ (31/2^{+})  \rightarrow	 29/2^{+} $&  11138.4(7) &  10200.0(4)         &   $<$ 0.2$^*$  &  -		     & -	    & M1 \\
941.0(1)  &$	15/2^{+}  \rightarrow	 13/2^{+} $&   5147.3(3) &  4206.3(2)          &   5.20(12)      &  0.65(7)$^{\$}$    & -0.01(1)     & M1 \\
947.7(1)  &$	 7/2^{-}  \rightarrow	  5/2^{-} $&   1014.9(1) &    67.2(1)          &   68.5(14)      &  1.38(4)$^{\&}$    & -0.04(1)      & M1+E2 \\
955.2(2)  &$	 9/2^{-}  \rightarrow	  7/2^{-} $&   2409.4(1) &  1454.4(1)          &   0.21(1)      &  -		     & -	    & M1 \\
967.4(7)  &$ 15/2^{(-)}  \rightarrow	 13/2^{+} $&   4999.3(2) &  4032.1(2)          &   $<$ 0.2$^*$  &  -		     & -	    & E1 \\
972.2(1)  &$	 9/2^{-}  \rightarrow	  7/2^{-} $&   1987.0(1) &  1014.9(1)          &   5.67(12)      &  0.43(2)$^{\$}$    & -0.02(1)      & M1+E2 \\
1014.9(1) &$	 7/2^{-}  \rightarrow	  3/2^{-} $&   1014.9(1) &  0.0(-)             &   31.8(6)      &  0.83(3)$^{\$}$    & 0.07(1)       & E2 \\
1042.0(1) &$   15/2^{+}  \rightarrow	 13/2^{+} $&   4476.9(2) &  3434.9(2)          &   13.2(3)      &  0.78(2)$^{\$}$    & -0.07(1)      & M1+E2 \\
1045.4(3) &$	21/2^{+}  \rightarrow	 19/2^{+} $&   7237.9(3) &  6192.9(2)          &   $<$ 0.2$^*$  &  -		     & -	    & M1 \\
1079.2(2) &$	 9/2^{-}  \rightarrow	  5/2^{-} $&   1987.0(1) &  907.6(1)           &   10.51(19)     &  1.01(3)$^{\$}$    & -0.02(2)      & E2 \\
1084.6(2) &$	27/2^{+}  \rightarrow	 25/2^{+} $&   9747.6(4) &  8663.0(3)          &   1.59(6)      &  -		     & -	    & M1 \\
1106.1(1) &$	 9/2^{+}  \rightarrow	  7/2^{-} $&   2121.0(1) &  1014.9(1)          &   68.4(2)      &  0.62(2)$^{\$}$    & 0.04(1)       & E1 \\
1113.4(1) &$	11/2^{-}  \rightarrow	  7/2^{-} $&   2128.3(1) &  1014.9(1)          &   35.7(7)      &  0.95(3)$^{\$}$    & 0.04(1)       & E2 \\
1128.2(5) &$	15/2^{+}  \rightarrow	 15/2^{+} $&   5147.3(3) &  4018.7(2)          &   1.26(5)      &  1.04(6)$^{\&}$    & - 	    & M1 \\
1132.3(5) &$	 5/2^{-}  \rightarrow	  3/2^{-} $&   1132.3(5) &  0.0(-)             &   $<$ 0.2$^*$  &  -		     & -	    & M1 \\
1136.8(1) &$	17/2^{+}  \rightarrow	 15/2^{+} $&   5155.7(2) &  4018.7(2)          &   1.42(5)      &  0.88(5)$^{\&}$    & - 	    & M1+E2 \\
1177.4(5) &$	11/2^{+}  \rightarrow	  9/2^{+} $&   3298.4(5) &  2121.0(1)          &   17.6(6)      &  1.24(9)$^{\&}$    & -0.03(1)      & M1+E2 \\
1178.6(5) &$   15/2^{+}  \rightarrow	 11/2^{+} $&   4476.9(2) &  3298.4(5)          &   5.3(9)       &  -		     & -	    & E2 \\
1223.6(2) &$ 15/2^{(+)}  \rightarrow	 11/2^{+} $&   4522.0(5) &  3298.4(5)          &   1.59(5)      &  1.78(10)$^{\&}$    & - 	    & (E2) \\
1246.8(2) &$	19/2^{+}  \rightarrow	 17/2^{+} $&   6065.1(2) &  4818.3(2)          &   1.08(2)      &  0.96(5)$^{\&}$    & -0.19(8)      & M1 \\
1271.2(1) &$ 13/2^{(-)}  \rightarrow	  9/2^{-} $&   3258.2(2) & 1987.0(1)           &   0.57(1)      &  0.95(6)$^{\$}$    & -		    & (E2) \\
1277.2(6) &$	 9/2^{-}  \rightarrow	  5/2^{-} $&   2409.4(1) & 1132.3(5)           &   $<$ 0.2$^*$  &  -		     & -	    & E2 \\
1291.8(3) &$	17/2^{+}  \rightarrow	 15/2^{+} $&   5310.7(2) & 4018.7(2)           &   4.31(12)     &  1.12(6)$^{\&}$    & -0.05(1)      & M1+E2 \\
1295.4(4) &$ (29/2^{+})  \rightarrow	 27/2^{+} $&  11043.0(6) &  9747.6(4)          &   $<$ 0.2$^*$  &  -		     & -	    & (M1) \\
1297.9(3) &$	13/2^{-}  \rightarrow	 11/2^{-} $&   3426.0(2) & 2128.3(1)           &   9.38(17)     &  0.59(4)$^{\$}$    & - 	    & M1 \\
1306.8(3) &$	13/2^{+}  \rightarrow	 11/2^{-} $&   3434.9(2) & 2128.3(1)           &   1.69(7)      &  0.44(2)$^{\$}$    & - 	    & E1 \\
1313.9(1) &$	13/2^{+}  \rightarrow	  9/2^{+} $&   3434.9(2) & 2121.0(1)           &   37.3(7)      &  0.99(3)$^{\$}$    & 0.05(1)       & E2 \\
1339.7(2) &$ 15/2^{(-)}  \rightarrow	 13/2^{-} $&   4766.0(3) & 3426.0(2)           &   1.17(7)      &  0.54(4)$^{\$}$    & -		    & (M1) \\
1341.0(5) &$	21/2^{+}  \rightarrow	 17/2^{+} $&   6971.6(3) & 5630.9(3)           &   $<$ 0.2$^*$  &  -		     & -	    & E2 \\
1341.8(3) &$	21/2^{+}  \rightarrow	 17/2^{+} $&   7266.9(2) & 5925.2(2)           &   1.76(4)      &  1.04(6)$^{\$}$    & 0.03(1)	    & E2 \\
1374.6(2) &$	19/2^{+}  \rightarrow	 17/2^{+} $&   6192.9(2) & 4818.3(2)           &   0.34(3)      &  -		     & -	    & M1 \\
1383.4(1) &$	17/2^{+}  \rightarrow	 13/2^{+} $&   4818.3(2) & 3434.9(2)           &   6.9(7)       &  0.93(5)$^{\$}$    & - 	    & E2 \\
1387.2(5) &$	 7/2^{-}  \rightarrow	  5/2^{-} $&   1454.4(1) & 67.2(1)             &   $<$ 0.2$^*$  &  0.66(5)$^{\$}$    & -		    & M1 \\
1394.2(2) &$	 9/2^{-}  \rightarrow	  7/2^{-} $&   2409.4(1) & 1014.9(1)           &   1.30(3)      &  -		     & - 	    & M1 \\
1412.0(2) &$	23/2^{+}  \rightarrow	 19/2^{+} $&   7604.9(3) &  6192.9(2)          &   0.65(2)      &  -		     & -	    & E2 \\
1438.7(2) &$	13/2^{-}  \rightarrow	  9/2^{-} $&   3426.0(2) &  1987.0(1)          &   3.75(8)      &  1.01(5)$^{\$}$    & 0.06(1)	    & E2 \\
1450.9(1) &$ 13/2^{(-)}  \rightarrow	  9/2^{-} $&   3258.2(2) &  1807.4(1)          &   0.17(1)      &  -		     & -	    & (E2) \\
1454.4(1) &$	 7/2^{-}  \rightarrow	  3/2^{-} $&   1454.4(1) &  0.0(-)             &   14.9(3)      &  0.96(3)$^{\$}$    & 0.10(6)      & E2 \\
1501.9(3) &$	27/2^{+}  \rightarrow	 23/2^{+} $&   9106.8(4) &  7604.9(3)          &   2.40(2)      &  1.57(8)$^{\&}$    & -		    & E2 \\
1505.1(3) &$ 13/2^{(+)}  \rightarrow	 13/2^{+} $&   4939.8(2) & 3434.9(2)           &   3.90(8)      &  0.96(5)$^{\$}$    & 0.01(2)      & (M1+E2) \\
1507.7(3) &$ 15/2^{(-)}  \rightarrow  13/2^{(-)}  $&    4766.0(3)&   3258.2(2)         &   $<$ 0.2$^*$  &  -		     & -		    & (M1) \\
1539.8(1) &$	23/2^{+}  \rightarrow	 19/2^{+} $&   7604.9(3) &   6065.1(2)         &   4.78(8)      &  0.96(6)$^{\$}$    & 0.02(1)	    & E2 \\
1578.6(1) &$	21/2^{+}  \rightarrow	 17/2^{+} $&   6734.3(2) &  5155.7(2)          &   5.36(11)      &  0.95(5)$^{\$}$    & 0.06(3)	    & E2 \\
1588.2(2) &$	19/2^{+}  \rightarrow    15/2^{+} $&   6065.1(2) &  4476.9(2)          &   4.39(10)	&  1.05(6)$^{\$}$    & 0.09(6)       & E2 \\
1607.0(3) &$	21/2^{+}  \rightarrow	 17/2^{+} $&   7237.9(3) & 5630.9(3)           &   $<$ 0.2$^*$  &  -		     & -	    & E2 \\
1611.8(1) &$	17/2^{+}  \rightarrow	 15/2^{+} $&   5630.9(3) &   4018.7(2          &   0.57(3)      &  -		     & -	    & M1 \\
1618.6(1) &$	13/2^{-}  \rightarrow	  9/2^{-} $&   3426.0(2) &  1807.4(1)          &   1.14(3)      &  1.85(9)$^{\&}$    & 0.03(1)	    & E2 \\
1683.9(1) &$	25/2^{+}  \rightarrow	 21/2^{+} $&   8950.8(2) &  7266.9(2)          &   0.66(4)      &  1.72(10)$^{\&}$    & 0.08(3)	    & E2 \\
1697.5(1) &$ 17/2^{(+)}  \rightarrow	 13/2^{+} $&   5729.6(2) & 4032.1(2)           &   0.45(2)      &  0.95(7)$^{\$}$    & -	            & (E2) \\
1712.2(3) &$	15/2^{+}  \rightarrow	 13/2^{+} $&   5147.3(3) & 3434.9(2)           &   0.55(3)      &  -		     & -	    & M1 \\
1720.8(1) &$	17/2^{+}  \rightarrow	 13/2^{+} $&   5155.7(2) &  3434.9(2)          &   12.6(2)      &  1.03(3)$^{\$}$    & 0.04(1)      & E2 \\
1740.2(1) &$	 9/2^{-}  \rightarrow	  5/2^{-} $&   1807.4(1) &   67.2(1)           &   15.4(2)      &  1.62(8)$^{\&}$    & 0.04(1)      & E2 \\
1753.7(2) &$	21/2^{+}  \rightarrow	 17/2^{+} $&   7678.9(2) &  5925.2(2)          &   $<$ 0.2$^*$  &  -                & -            & E2 \\
1788.0(3) &$	13/2^{-}  \rightarrow	  9/2^{-} $&   4197.1(2) &  2409.4(1)          &   0.84(4)      &  -		     & -	    & E2 \\
1822.0(3) &$ (31/2^{+})  \rightarrow	 27/2^{+} $&  10928.8(5) & 9106.8(4)           &   1.37(5)      &  -		     & -	    & (E2) \\
1835.0(5) &$	23/2^{+}  \rightarrow	 19/2^{+} $&   8027.7(2) &  6192.9(2)          &   $<$ 0.2$^*$  &  -		     & -	    & E2 \\
1848.6(2) &$	15/2^{+}  \rightarrow	 11/2^{+} $&   5147.3(3) & 3298.4(5)           &   1.43(5)      &  1.50(8)$^{\&}$    & - 	    & E2 \\
1875.8(1) &$	17/2^{+}  \rightarrow	 13/2^{+} $&   5310.7(2) & 3434.9(2)           &   1.12(5)      &  1.13(7)$^{\$}$    & -		    & E2 \\
1903.8(1) &$	13/2^{+}  \rightarrow	 11/2^{-} $&   4032.1(2) & 2128.3(1)           &   2.25(5)      &  0.75(8)$^{\$}$    & 0.06(2) 	    & E1 \\
1919.8(1) &$	 9/2^{-}  \rightarrow	  5/2^{-} $&   1987.0(1) &   67.2(1)           &   15.2(2)      &  1.07(3)$^{\$}$    & 0.07(2)      & E2 \\
1945.4(7) &$ (35/2^{+})  \rightarrow  (31/2^{+})  $&  12874.2(9) &   10928.8(5)        &   0.79(2)      &  1.71(9)$^{\&}$    & -	            & (E2) \\
1962.8(5) &$	23/2^{+}  \rightarrow	 19/2^{+} $&   8027.7(2) &  6065.1(2)          &   $<$ 0.2$^*$  &  -		     & -	    & E2 \\
2042.4(5) &$ (29/2^{+})  \rightarrow	 25/2^{+} $&  10993.2(6) &  8950.8(2)          &   $<$ 0.2$^*$  &  -		     & -	    & (E2) \\
2044.9(3) &$	13/2^{+}  \rightarrow	  9/2^{-} $&   4032.1(2) & 1987.0(1)           &   0.86(15)     &  0.97(7)$^{\$}$    & - 	    & M2 \\
2048.0(4) &$	21/2^{+}  \rightarrow	 17/2^{+} $&   7678.9(2) & 5630.9(3)           &   $<$ 0.2$^*$  &  -		     & -	    & E2 \\
2068.8(2) &$	13/2^{-}  \rightarrow	 11/2^{-} $&   4197.1(2) & 2128.3(1)           &   2.62(22)     &  0.46(3)$^{\$}$    & - 	    & M1+E2 \\
2078.0(2) &$	13/2^{+}  \rightarrow	 11/2^{-} $&   4206.3(2) & 2128.3(1)           &   5.3(7)       &  0.69(5)$^{\$}$    & - 	    & E1 \\
2078.8(2) &$   19/2^{+}  \rightarrow    15/2^{+}  $&   6555.7(3) & 4476.9(2)           &   -$^{\#}$	 & -               & -	            & E2 \\
2082.2(2) &$	21/2^{+}  \rightarrow	 17/2^{+} $&  7237.9(3)  &  5155.7(2)          &   - $^{\#}$	&  0.86(5)$^{\$}$    & 0.14(5)      & E2 \\
2085.3(2) &$	13/2^{+}  \rightarrow	  9/2^{+} $&   4206.3(2) & 2121.0(1)           &   1.5(6)       &  1.47(8)$^{\&}$    & - 	    & E2 \\
2111.2(1) &$	21/2^{+}  \rightarrow	 17/2^{+} $&   7266.9(2) &  5155.7(2)          &   1.4(6)       &  0.86(5)$^{\$}$    & 0.16(9)	    & E2 \\
2196.0(2) &$	17/2^{+}  \rightarrow	 13/2^{+} $&   5630.9(3) & 3434.9(2)           &   0.66(4)      &  0.88(5)$^{\$}$    &		    & E2 \\
2216.6(2) &$	25/2^{+}  \rightarrow	 21/2^{+} $&   8950.8(2) &   6734.3(2)         &   1.45(11)     &  1.25(9)$^{\$}$    & - 	    & E2 \\
2224.8(1) &$	13/2^{+}  \rightarrow	  9/2^{-} $&   4032.1(2) &  1807.4(1)          &   0.85(12)     &  -		     & -	    & M2 \\
2342.0(1) &$	 9/2^{-}  \rightarrow	  5/2^{-} $&   2409.4(1) &   67.2(1)           &   2.19(4)      &  -		     & -	    & E2 \\
2368.2(2) &$	21/2^{+}  \rightarrow	 17/2^{+} $&   7678.9(2) &  5310.7(2)          &   0.23(2)      &  -		     & - 	    & E2 \\
2490.3(1) &$	17/2^{+}  \rightarrow	 13/2^{+} $&   5925.2(2) &  3434.9(2)          &   1.21(4)      &  1.11(7)$^{\$}$    & 0.02(1)	    & E2 \\
2523.2(4) &$	21/2^{+}  \rightarrow	 17/2^{+} $&   7678.9(2) &   5155.7(2)         &   0.80(8)      &  1.16(8)$^{\$}$    & - 	    & E2 \\
2537.0(8) &$   19/2^{+}  \rightarrow	 15/2^{+} $&   6555.7(3) &  4018.7(2)          &   0.42(13)     &  1.64(9)$^{\&}$    & - 	    & E2 \\
2634.6(1) &$	13/2^{+}  \rightarrow	 11/2^{-} $&   4762.9(2) &  2128.3(1)          &   2.75(6)      &  0.73(4)$^{\$}$    & 0.12(7)	    & E1 \\
2818.8(1) &$ 13/2^{(+)}  \rightarrow	  9/2^{+} $&   4939.8(2) &  2121.0(1)          &   1.06(5)      &  1.62(9)$^{\&}$    & - 	    & (E2) \\
2860.6(2) &$	21/2^{+}  \rightarrow	 17/2^{+} $&   7678.9(2) &  4818.3(2)          &   0.27(1)      &  1.29(11)$^{\$}$    & -		    & E2 \\
2871.0(2) &$ 15/2^{(-)}  \rightarrow	 11/2^{-} $&   4999.3(2) &  2128.3(1)          &   1.50(5)      &  0.83(5)$^{\$}$    & - 	    & (E2) \\
3035.6(1) &$ (15/2^{-})  \rightarrow	 11/2^{-} $&   5163.9(2) &  2128.3(1)          &   1.88(6)      &  -		     & -	& (E2) \\
  
\hline 

\label{int-61Ni}
\footnotetext[1]{Relative $\gamma$-ray intensities are estimated from prompt spectra and\\
normalized to 100 for the total intensity of 67-keV $\gamma$-ray.}\\
\footnotetext[2]{Unobserved transition, Adopted from Ref~\cite{Saradindu}}
\footnotetext[3]{From 1314~keV (E2) DCO gate;}
\footnotetext[4]{From 1113~keV (E2) DCO gate;}
\footnotetext[5]{From 1920~keV (E2) DCO gate;}
\footnotetext[6]{From 1439~keV (E2) DCO gate;}
\footnotetext[7]{From 1106~keV (E1) DCO gate;}
\footnotetext[8]{From 1721~keV (E2) DCO gate;}
\footnotetext[9]{From 593~keV (E1) DCO gate;}

\footnotetext[11]{Adopted from Ref ~\cite{Saradindu} or ~\cite{Saradindu}}

\end{longtable*}

\end{document}